\documentclass[manuscript,screen,9pt]{acmart}

\AtBeginDocument{%
  \providecommand\BibTeX{{%
    \normalfont B\kern-0.5em{\scshape i\kern-0.25em b}\kern-0.8em\TeX}}}

\setcopyright{acmcopyright}
\copyrightyear{2018}
\acmYear{2018}
\acmDOI{XXXXXXX.XXXXXXX}

\acmConference[Conference acronym 'XX]{Make sure to enter the correct
  conference title from your rights confirmation emai}{June 03--05,
  2018}{Woodstock, NY}
\acmPrice{15.00}
\acmISBN{978-1-4503-XXXX-X/18/06}

\usepackage{comment}

\usepackage{color}
\usepackage{ulem}
\usepackage{graphicx}
\usepackage{subfig}
\usepackage{algorithmic}
\usepackage{algorithm}
\usepackage{amsfonts}
\usepackage{tikz}
\usepackage{xspace}

\usepackage{setspace}

\usepackage{wrapfig}

\newcommand{\snote}[1]{{\color{blue} #1}}
\definecolor{darkgreen}{rgb}{0.0,0.5,0.0}
\newcommand{\arya}[1]{{\color{darkgreen} #1}}

\newcommand{\cvf}{\textit{cvf}\xspace}
\newcommand{\feasiblecvf}{\ensuremath{cvf_{feasible}}\xspace}
\newcommand{\maxcvf}{\ensuremath{cvf_{max}}\xspace}
\newcommand{\mrank}{\textit{max-rank}\xspace}
\newcommand{\arank}{\textit{average-rank}\xspace}
\newcommand{\transeff}{\ensuremath{effect_{prog}}\xspace}
\newcommand{\cvfeff}{\ensuremath{effect_{cvf}}\xspace}
\newcommand{\cvfintv}{\textit{cvf\_interval}\xspace}
\newcommand{\convstep}{\textit{convergence\_steps}\xspace}
\newcommand{\relcvf}{\ensuremath{rel_{cvf}}\xspace}

\newtheorem{observation}{Observation}

\begin{document}

\title{Technical Report: Using Static Analysis to Compute Benefit of Tolerating Consistency Violation Faults}

\author{Duong Nguyen}
\email{dn347@georgetown.edu}
\affiliation{%
  \institution{Computer Science, Georgetown University}
  \city{Washington DC}
  \country{USA}
  \postcode{20057}
}

\author{Arya Tanmay Gupta}
\email{atgupta@msu.edu}
\affiliation{%
  \institution{Computer Science and Engineering, Michigan State University}
  \city{East Lansing}
  \country{USA}
  \postcode{48825}
}

\author{Sandeep S Kulkarni}
\email{sandeep@msu.edu}
\affiliation{%
  \institution{Computer Science and Engineering, Michigan State University}
  \city{East Lansing}
  \country{USA}
  \postcode{48825}
}


\renewcommand{\shortauthors}{Nguyen et al.}

\begin{abstract}
Synchronization is the Achilles heel of concurrent programs. Synchronization requirement is often used to ensure that the execution of the concurrent program can be serialized. Without synchronization requirement, a program suffers from consistency violations. Recently, it was shown that if programs are designed to tolerate such consistency violation faults (\cvf{s}) then one can obtain substantial performance gain. Previous efforts to analyze the effect of \cvf-tolerance are limited to run-time analysis of the program to determine if tolerating \cvf{s} can improve the performance. Such run-time analysis is very expensive and provides limited insight. 

In this work, we consider the question, `Can static analysis of the program predict the benefit of \cvf-tolerance?' We find that the answer to this question is affirmative. Specifically, we use static analysis to evaluate the cost of a \cvf and demonstrate that it can be used to predict the benefit of \cvf-tolerance. We also find that when faced with a large state space, partial analysis of the state space (via sampling) also provides the required information to predict the benefit of \cvf-tolerance. Furthermore, we observe that the \cvf-cost distribution is exponential in nature, i.e., the probability that a \cvf has a cost of $c$ is $A.B^{-c}$, where $A$ and $B$ are constants 
, i.e., most \cvf{s} cause no/low perturbation whereas a small number of \cvf{s} cause a large perturbation. This opens up new aveneus to evaluate the benefit of \cvf-tolerance.

\end{abstract}


\maketitle


\section{
Introduction}

As we see an end to Moore's law, parallel/concurrent computing has become essential to solving problems efficiently because now, instead of increasing the power of processors, technology is witnessing integration of multiple processors on the same chip. While an individual processor becomes faster over time, this increased speedup is dwarfed by the increase in the number of processors/cores.
To fairly utilize such an architecture now, mutiprocess computing is implemented. In such implementations, the task is to solve the same problem collectively; processes share data and communicate with each other to achieve so. Such implementations is done in two ways based on how far processors are placed from each other. If the processors are placed on a single chip, then the processes communicate with one another under the shared memory model. If, on the other hand, the processors are placed remotely from each other, then the processes communicate with one another under the message passing model.

In multiprocessor computing, it is always desired that the system implemented is \textit{execution faithful}, that is the execution of the program appears as if it was executed by a uniprocess program. 
To achieve this, programs are often implemented under synchronization constraints, e.g., locks, locksteps, and semaphores. If synchronization requirements are violated then a process would read inconsistent values of variables. 



The authors of
\cite{NKD2019ICDCN.short} introduced the notion of consistency violation faults (\cvf{s}) to model such improper simultaneous access to shared data. For example, in case of the vertex cover problem, simultaneous updates may cause two adjacent nodes to enter the vertex cover simultaneously. The \cvf{s} could cause the program to diverge from its goal state (e.g., a minimal vertex cover). However, the underlying program is able to handle such inconsistency to reach the goal state. In \cite{NK2020SRDS.short},
authors have considered various problems including token ring, graph coloring, and maximal matching.
Here, it was observed that there is a substantial benefit in permitting \cvf{s} and tolerating them rather than eliminating them using synchronization approaches such as local mutual exclusion.

Now after establishing that tolerating \textit{cvf}s is beneficial, especially in terms of convergence time, it is desired that we are able to predict the overheads involved in tolerating the \textit{cvf}s, especially when the problem size increases.
Currently, the only known approach to make such predictions is experimental analysis where you consider the execution of the program without \cvf{s} (i.e., with suitable synchronization) and with \cvf{s} (where \cvf{s} are permitted). It is highly desirable that we can predict the expected improvement from \cvf tolerance without such experimental analysis which is both time consuming and does not allow one to predict the expected benefit under different execution conditions (e.g., level of concurrency).

One way to ameliorate this is to predict the expected performance gain via static analysis. If feasible, this would indicate that we can analyze the given program once so we can identify the expected benefit of the given program by simply analyzing its code rather than by run-time analysis. In turn, this would allow us to predict the expected performance gain under different 
levels of concurrency and different memory models.


With this motivation, the \textit{first} goal of this work is to understand if static analysis of programs can provide sufficient information about the ability to gain performance by tolerating \cvf{s} instead of eliminating them. 

It is also important to address known limitations of static analysis, viz. state space explosion. Specifically, as problem size increases, the size of the state space increases exponentially thereby making static analysis difficult/impossible. The way to get around this is to generalize from small programs and to do sampling of state space. If successful, we can get an understanding of how \cvf-tolerance will improve performance of large programs using small programs. Or, we can only partially analyze the state space to identify expected performance gain from partial analysis of the state space.

With this motivation, the \textit{second} goal of this work is to understand if partial analysis of state space provides sufficient information to gauge the performance gain provided by full static analysis. Or, if the performance gain of small programs can be generalized to the corresponding large programs. 


The \textit{third} goal of this paper is to determine how the cost of \cvf{s} varies in different programs. Specifically, if we can empirically determine the distribution cost of \cvf{s} then it would allow us to benefit from existing tools for statistical analysis. 

Based on the above discussion, we focus on static analysis of three programs, for distributed token ring, graph coloring and matching on message passing model. These problems were chosen because they are self-stabilizing \cite{EDW426} which implies that they are able to tolerate \cvf{s} (cf. Section \ref{sec:cvf} for justification). For these programs, we analyze their state space and define a rank of each state which captures how long it will take for the program to reach the goal state (e.g., valid coloring) from that state. Then, we can define the benefit of a regular program transition (change in rank due to a program transition) and the cost of a \cvf (change in rank by a \cvf). Subsequently, we analyze how the benefit of program and \cvf transitions varies across programs and input sizes. 

The contributions of the paper are as follows:
\begin{itemize}
\item 
Regarding our first goal, we show that static analysis, indeed, provides sufficient information about expected speedup. 
\item 
Regarding our second goal, we show that partial analysis of the state space does provide sufficient information about the performance in the presence of \cvf{s}. In turn, this indicates that one can generalize results from smaller programs to larger programs (with more number of nodes/edges).  
\item 
Regarding the third goal,  we find that the cost distribution of \cvf{s} is exponential in nature. In other words, the probability that a given \cvf perturbs the rank by $c$ is $AB^{-c}$ where $A$ and $B$ are constants.
This means that most \cvf{s} have low cost whereas a few \cvf{s} have a large cost. In turn, this implies that existing statistical techniques can provide information about performance gain in the presence of \cvf{s}.

\item Additionally, we analyze the cost distribution of \cvf{s} in these programs. We compute the trend of the same program; this trend predicts the cost distribution of \cvf{s} as the problem size increases.
\item By focusing on the average benefit of program transitions and \cvf{s} (under the assumption that they are equally likely), we compute $\relcvf$ which denotes the average number of program transitions needed to undo the effect of a \cvf. We compute $\relcvf$ for our case studies using full analysis (where we construct the entire state space to determine ranks of each states) and partial analysis (where rank computation is done by sampling). We consider \arank and \mrank. We find that partial analysis provides results that are close to the full analysis thereby demonstrating that we can consider only sampling based approach when the state space is large.

\end{itemize}

\textbf{Organization of the paper. }
The rest of the paper is organized as follows. In Section \ref{sec:modeling}, we define the computational model. We introduce \cvf{s} in Section \ref{sec:cvf}. Section \ref{sec:modelcost} presents our approach for computing the cost of \cvf{s}. Section \ref{sec:computecostbenefit} identifies the cost distribution of \cvf{s}.  In Section \ref{sec:analyzerelativecost}, we compute the relative cost of \cvf{s} compared to the benefit of program transitions. We evaluate the steps required for convergence in the presence of \cvf{s} in Section \ref{sec:performance}. In Section \ref{sec:interpret}, we interpret these results and discuss their implications for executing concurrent algorithms. Section \ref{sec:related} discusses related work. Finally, Section \ref{sec:concl} provides conclusion
.

\section{Modeling Parallel/Concurrent Programs}
\label{sec:modeling}

A distributed/parallel/concurrent program consists of a set of $n$ processes \cite{ag94}.
Each process $j$ has a set of neighboring processes, denoted as $Nb.j$. Each process $j$ is associated with a set of variables and a set of actions.
Each action of process $j$ is of the form $g \longrightarrow st$, where $g$ is a Boolean expression involving variables of $j$ and the variables of its neighbors $(Nb.j)$, and $st$ is the statement that updates the variables of $j$. 

The state of a process $j$ is obtained by assigning values to all the variables of $j$ from their respective domains. Variables of program $p$ are the union of the variables of its processes. A state of  program $p$ is obtained by assigning values to all the variables of $p$. The state space $S_p$ of program $p$ is the set of all possible states of $p$.

We say that an action $g\longrightarrow st$ of some process is enabled in a state $s$ iff $g$ evaluates to true in $s$. Transitions of process $j$ in program $p$ is the set $\{(s_0, s_1) | s_0, s_1 \in S_p$ and $s_1$ is obtained from $s_0$ by executing one of the enabled actions of $j\}$. The transition of program $p$ is the union of the transitions of all its processes.

A computation of program $p$ is of the form $\langle s_0, s_1, \dots \rangle$, where $s_i, i \geq 0$ is a state of the program and $(s_i,s_{i+1})$ is a transition of $p$. For the sake of simplicity, we assume that each state has an outgoing transition, i.e., for any state $s$, there exists a state $s'$ such that $(s,s')$ is a transition of $p$. To achieve this, if no action in any process is enabled in state $s$ then  we add $(s, s)$ as a transition of $p$. 

A state predicate $S$ is a Boolean expression involving variables of $p$. We also overload $S$ to denote a subset of the state space where 
predicate $S$ is true. In other words, we use the Boolean expression $x=0$ and the set of all states where the $x$ value is $0$ interchangeably. 

\section{Consistency Violation Faults.}
\label{sec:cvf}

The program computation model assumes that each action execution is atomic. Thus, a consistency violation happens when a process $j$ reads old values of one or more other processes, and performs an execution based on those values. 
Reading old values may cause process $j$ to incorrectly go from state $s_0$ to $s_1$ where $(s_0, s_1)$ is not a transition of $j$. We denote such execution as a \textit{consistency violation fault} (\cvf).


We illustrate \cvf{s} with an example of a simple graph coloring program where a node changes its color to another available color if it finds that its color matches with that of its neighbor (see Appendix~\ref{sec:case-study} for detailed descriptions of the program).  Consider the state $\langle c.0\!=\!0,c.1\!=\!0\rangle$ of this program where processes $0$ and $1$ are neighbors. 
If process $0$ (respectively $1$) executes then the resulting state would be $\langle c.0\!=\!1,c.1\!=\!0\rangle$ (respectively, $\langle c.0\!=\!0,c.1\!=\!1\rangle$). On the other hand, if both processes execute \textit{simultaneously} then the resulting state would be $\langle c.0\!=\!1,c.1\!=\!1\rangle$. Clearly, this is undesirable and the colors may remain invalid forever. 

This problem can be addressed via \textit{local mutual exclusion} \cite{10.1007/3-540-40026-5_15,KY2002SRDS,NA02JPDC,dima} that guarantees that two neighboring processes do not execute at the same time.
Clearly, local mutual exclusion introduces an overhead as it requires extra work to coordinate among the processes and block the waiting processes. 

While the above example is focused on \textit{simultaneous execution} of the program actions, such execution is an instance of \textit{consistency violation faults} where the process executing the action reads an incorrect value for one of its neighbors' variables. In this example, transition from $\langle c.0\!=\!0,c.1\!=\!0\rangle$ to $\langle c.1\!=\!1,c.1\!=\!1\rangle$ can be thought of as a transition from $\langle c.0\!=\!0,c.1\!=\!0\rangle$ to $\langle c.0\!=\!1,c.1\!=\!0\rangle$ by process $0$ and then transition from $\langle c.0\!=\!1,c.1\!=\!0\rangle$ to $\langle c.0\!=\!1,c.1\!=\!1\rangle$ by process $1$ where it reads an old value ($c.0\!=\!0$) of $c.0$. This latter transition is an instance of \cvf. 

It follows that consistency violation faults associated with program $p$ are subset of $S_p \times S_p$, where $S_p$ is the state space of $p$.  Hence, we can extend the definition of a computation as follows. 

\textbf{Computation in the presence of \cvf{s}. } 
A sequence $\langle s_0, s_1, s_2, \dots\rangle$ is a computation of program $p$ in the presence of $\cvf.p$ iff for each $i$, $(s_i, s_{i+1})$ is a transition of $p$ or is in $\cvf.p$. 
(Note that $\cvf{s}$ are not required to be terminating, {i.e. the number of \cvf{s} can be infinite}.)



Permitting and tolerating \cvf{s} differs from traditional approaches of concurrent computing where the goal is to keep the state of the program to be correct by ensuring that action execution can always be viewed to be atomic. As we have seen, \cvf{s} have the potential to perturb the program computation from its goal states. Our goal in this paper is to analyze the properties of such \cvf transitions. We study \cvf{s} in the graph coloring problem, token ring problem and maximal matching problem.
Specifically, we study the distribution of the \textit{cost} of \cvf{s} in these programs. 
We also evaluate the cost of these \cvf{s} while analyzing the execution time for these programs. 




\subsection{Self-stabilization and \cvf{s}}

In this section, we review the definition of self-stabilizing programs and relate it to \cvf{s}. A stabilizing program is guaranteed to recover from an arbitrary state to its legitimate states. Specifically,

Let $p$ be a program and $inv$ be a state predicate of $p$. We say that $p$ is self-stabilizing (respectively, stabilizing) with invariant $inv$ iff
\begin{itemize}
    \item Any program computation includes a state where $inv$ is true. In other words, for any computation $\langle s_0, s_1, \dots \rangle$, there exists a state $s_l$ such that $s_l \in inv$
    \item If $(s_0,s_1)$ is a transition of $p$ and $s_0 \in inv$ then $s_1 \in inv$. 
\end{itemize}


In other words, starting from an arbitrary state, a stabilizing program $p$ reaches a state in $inv$ and stays there forever. If $p$ is stabilizing to invariant $inv$ then $inv$ is the legitimate states of program $p$. 

We observe that a stabilizing program naturally tolerates (finite number of) \cvf{s}. In other words,

\begin{observation}
Let $p$ be a stabilizing program with invariant $inv$. Let $\sigma = s_0, s_1, \cdots$ be a computation of $p$ in the presence of \cvf{s}. If the number of occurrences of \cvf{s} is finite in $\sigma = s_0, s_1, \cdots$ then $\sigma$ will reach a state in $inv$ and stay there forever. 
\end{observation}

This follows from the fact that if \cvf{s} are finite then once they are done, the program is guaranteed to recover due to the property of stabilization. We focus on analyzing probabilistic stabilization \cite{Herman90IPL} of stabilizing programs when \cvf{s} are not finite. 


\section{Modeling Cost of \cvf{s}}
\label{sec:modelcost}

While determining the \cvf{s} for a given process/program is potentially a challenging task, we can see that \cvf{s} at $j$, namely $cvf.j$ is a subset of $maxcvf.j$, where 

$maxcvf.j =$ 
\{ $(s, s') | s'$ is obtained by changing variables of $j$ in $s$ \}

Similar to transitions of a program, we can define \cvf{s} for a program. In other words, $\cvf.p$ ($maxcvf.p$, respectively) for program $p$ is the union of $cvf.j$ ($maxcvf.j$) where $j$ is a process in $p$. 

\begin{figure}[ht]
    \centering
    \includegraphics[width=0.45\textwidth]{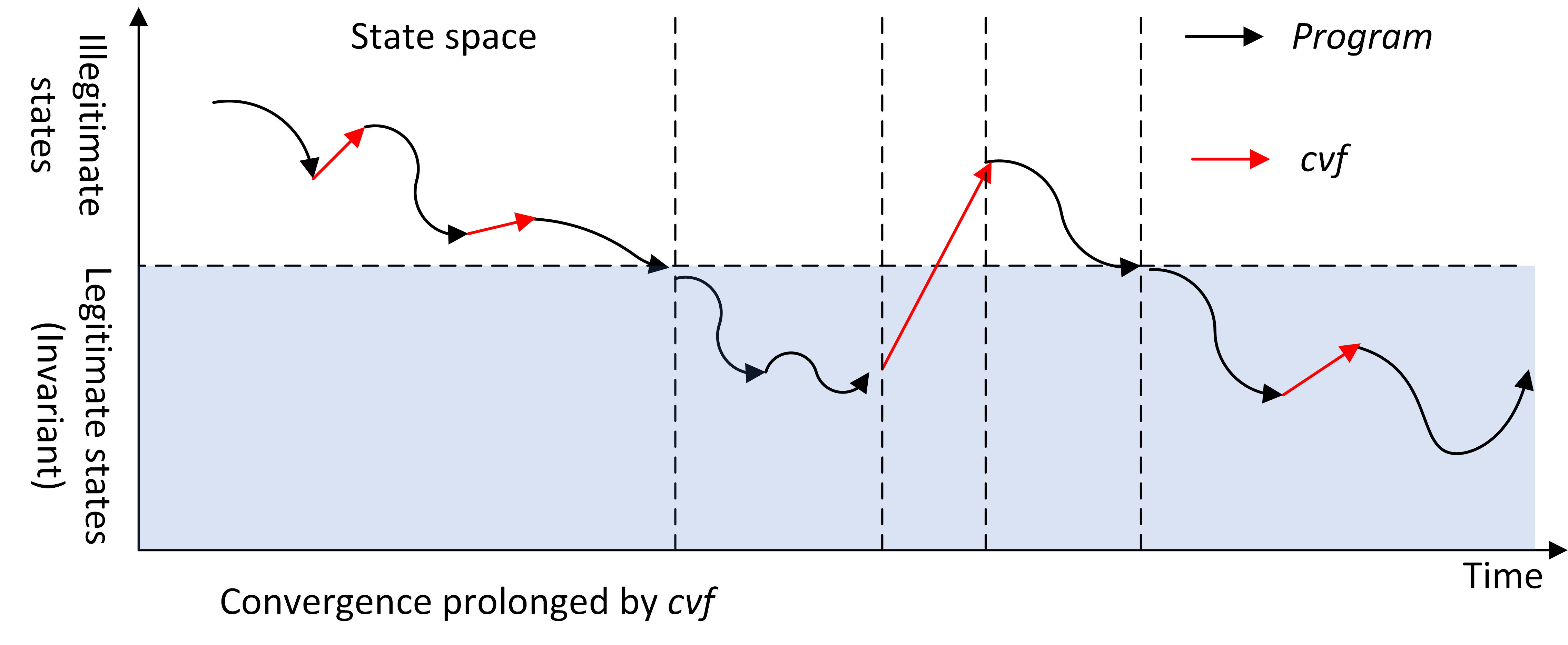}
    \Description{find me}
    \caption{Computation in the Presence of \cvf{s}}
    \label{fig:stabilizationillustration}
\end{figure}

Let  $\langle s_0, s_1, s_2, \dots\rangle$ be a computation of program $p$ in the presence of $\cvf.p$~.
Thus, we can view the execution of a stabilizing program as shown in Figure \ref{fig:stabilizationillustration}
where the rank of each state is depicted by how far the state is from a legitimate state (a state in the invariant).
Examples of such rank functions include (minimum, average, or maximum) steps/time required to reach a state in $inv$. 
Specifically, the rank of states in $inv$ is $0$ and a sufficiently long program computation decreases the rank of the program state. 

Note that execution of \cvf{s} \textit{may} cause the rank value to increase thereby increasing steps for recovery (or preventing the recovery).
However, the time required to execute one step is lower if \cvf{s} are permitted to occur, as there would be no need to add any synchronization to ensure atomicity of action execution. 



To guarantee convergence in spite of \cvf{s}, we need to ensure that the rank decreases in a computation over time even if it does not decrease in every step. 
In other words, consider a sufficiently long computation in the presence of \cvf{s}: $\langle s_0, s_1, \cdots, s_l, \dots\rangle$, where all states in the computation are outside legitimate states. 
In this case, we need to ensure that $rank(s_0) > rank(s_l)$. If this is assured then eventually the program will reach a state whose rank is $0$, thereby guaranteeing the recovery/valid termination of the program. 
Additionally, \cvf{s} are not malicious in nature since they are caused when a node reads an older value for one of its neighbors. 
In other words, \cvf{s} are random and unintentional. 
Hence, we can focus on the \textit{average} effect of \cvf (using statistical analysis) instead of the \textit{worst-case} effect. 
%
With this motivation, the goal of the paper is to study rank changes by program and \cvf transitions. 


\textbf{Average rank change by \cvf{s}. }
Since identifying precise \cvf transitions for a given program is a challenging task, we can approximate it with $maxcvf$. 
We have various choices for defining the rank of a state. For example, we could use the average/minimum/maximum number of transitions required to reach a legitimate state. 
One observation in this context is that the increase (respectively, decrease) in the rank from $s_0$ to $s_1$ $(rank(s_1)-rank(s_0))$  is the same as the decrease (respectively, increase) in the rank from $s_1$ to $s_0$. Moreover, if $(s_0,s_1) \in  \maxcvf$ then $(s_1,s_0) \in  \maxcvf$. Thus, if we average the change in rank of all transitions in  $\maxcvf$ then the average is $0$. 

\begin{observation}
The average change in rank by transitions in  $\maxcvf$ is $0$.
\end{observation}

What the above result shows is interesting and misleading. It is interesting because it shows that {if} all \cvf{s} were equally likely then the overall benefit and loss caused by \cvf{s} would be $0$. 
At the same time, this result is misleading because the actual \cvf{s} that could occur are a subset of those in $\maxcvf$. 
Since our goal is to determine the frequency of \cvf{s} under which convergence property is not affected, we only focus on \cvf{s} that increase the rank.

\textbf{Defining $\feasiblecvf$. } We also introduce the notion of $\feasiblecvf$. This is also a superset of actual \cvf{s} that could occur in the program but a subset of $\maxcvf$. To understand the notion of $\feasiblecvf$, observe that, in action $g \longrightarrow st$, the process may read arbitrary values of its neighbors but it will change its state by executing $st$. 
For example, if $st$ was of the form $x=x+1$ where $x$ is a local variable of the process then a \cvf cannot perturb the program from a state where $x=1$ to $x=0$ {(note that this perturbation is allowed in \maxcvf)}. 
Since \cvf{s} of a given program can be very difficult to compute, we use $\feasiblecvf$ as an approximation of \cvf{s}. Note that the average rank of transitions in $\feasiblecvf$ need not be $0$. Furthermore, we only consider those \cvf{s} that increase the rank.



\section{Computing the Benefit of Program Transitions and Cost of \cvf{s}}
\label{sec:computecostbenefit}

As discussed earlier, to compute the benefit of a program transition (respectively, cost of \cvf) $(s_0, s_1)$, we need to find the \textit{ranks} of $s_0$ and $s_1$ and the benefit of the program transition (respectively, cost of \cvf) is $rank(s_1)-rank(s_0)$. 
There are several choices to define the rank subject to the constraint that the rank of legitimate states should be $0$ and the rank of illegitimate states should be positive. We consider two options 

\begin{itemize}
    \item \textit{\mrank}, where $\mrank(s_0)$ is the length of the maximum path from $s_0$ to reach a legitimate state.  
    \item \textit{\arank}, where $\arank(s_0)$ is the average length of all the paths from $s_0$ to reach a state in the invariant. 
\end{itemize}

Note that for a stabilizing program, the rank will always be finite since each computation is eventually guaranteed to reach the invariant. Furthermore, we can utilize existing algorithms to compute \mrank and \arank. Outline of these algorithms is in Appendix \ref{sec:define-ranks}.
Once the rank of every state is computed, we compute the benefit of a program transition $(s_0, s_1)$ to be $rank(s_1)-rank(s_0)$. Then, we compute the average benefit of all program transitions. For \cvf{s}, as discussed above, we only consider \cvf{s} that disturb the convergence, i.e $\cvf = (s_0, s_1)$ where $rank(s_0) < rank(s_1)$. We compute the average cost of those \cvf{s}.

\subsection{Case Studies}

We consider three case studies: token ring program by Dijkstra \cite{EDW426}, Graph Coloring program by Gradinariu and Tixeuil~\cite{GT2000OPODIS.short} and maximal matching program by
Manne et al.~\cite{MMPT2009TCS.short}. Since the description of the algorithms is not critical for the subsequent discussion, for reasons of space, we describe them in Appendix. The only property we use of these algorithms is that they are stabilizing. 

\subsection{Program and Cvf distribution 
}

\begin{figure*}[ht]
    \vspace{-10pt}
    \centering
    \subfloat[]{
        \includegraphics[width=0.3\textwidth]{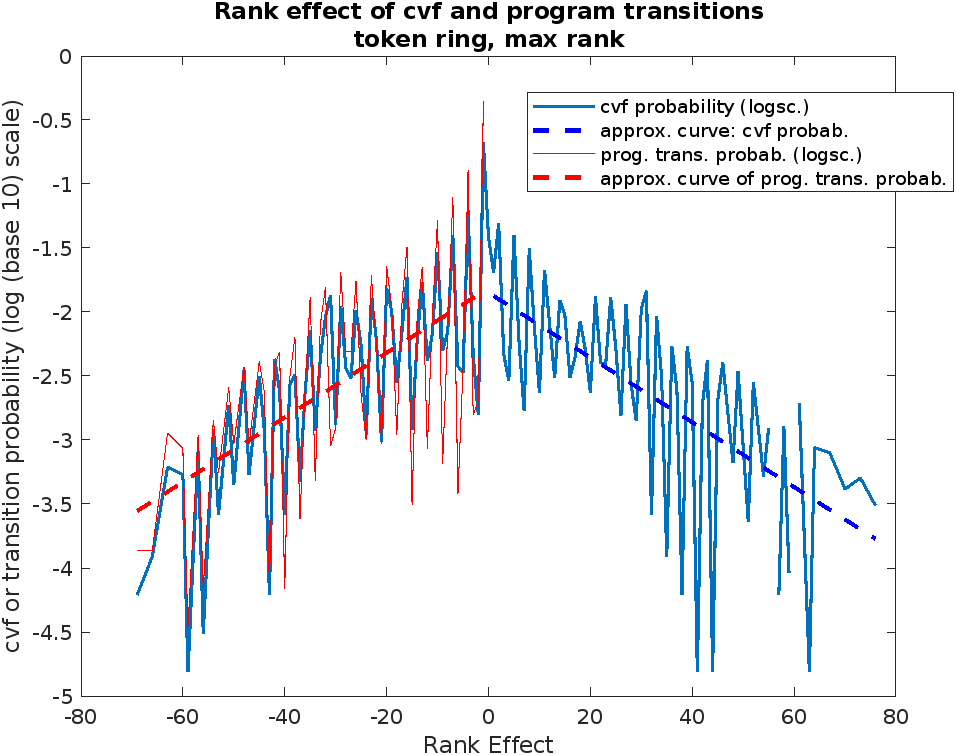}
        \label{fig:histogram-line-dijkstra-max}
    }
    \subfloat[]{
        \includegraphics[width=0.3\textwidth]{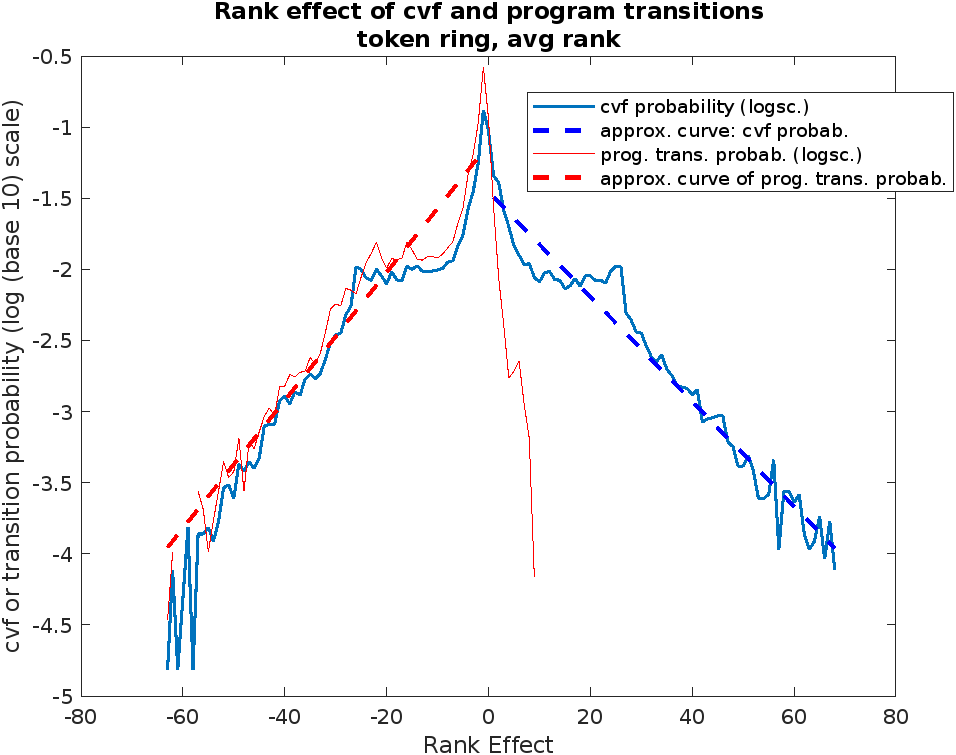}
        \label{fig:histogram-line-dijkstra-avg}
    }
    \subfloat[]{
        \includegraphics[width=0.3\textwidth]{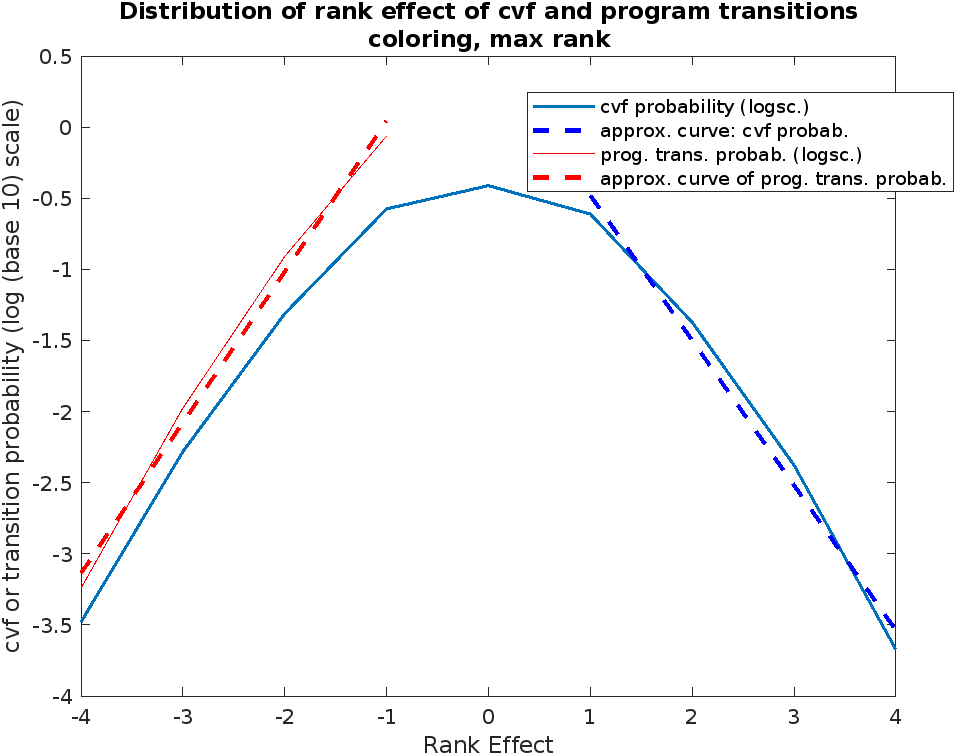}
        \label{fig:histogram-bar-coloring-max}
    }
    \\
    \subfloat[]{
        \includegraphics[width=0.3\textwidth]{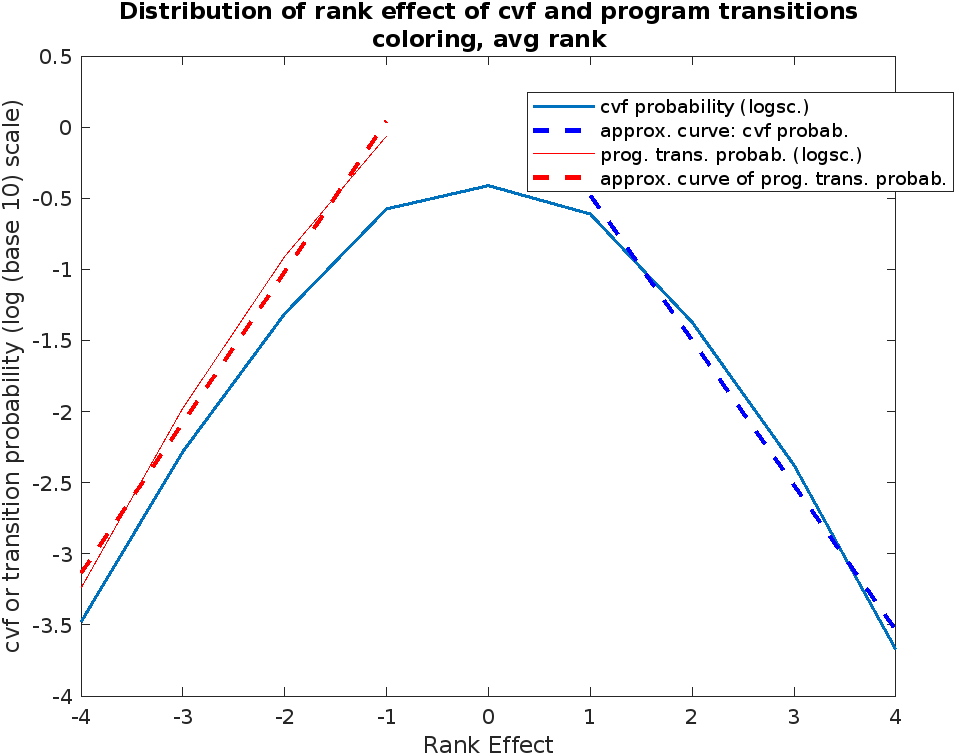}
        \label{fig:histogram-bar-coloring-avg}
    }
    \subfloat[]{
        \includegraphics[width=0.3\textwidth]{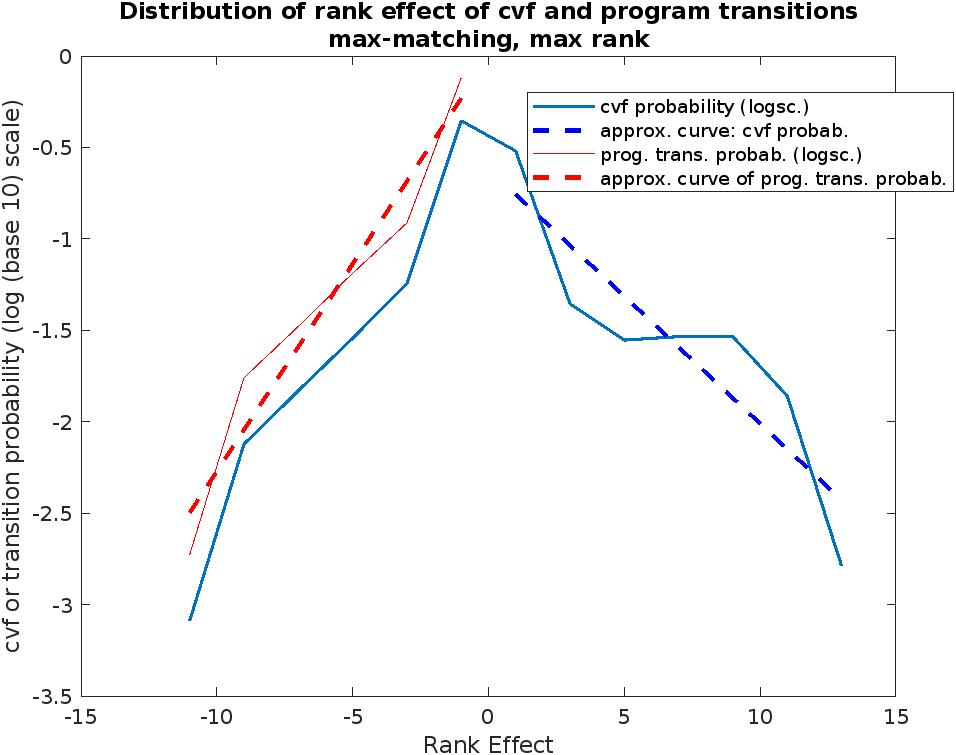}
        \label{fig:histogram-line-max-match-max}
    }
    \subfloat[]{
        \includegraphics[width=0.3\textwidth]{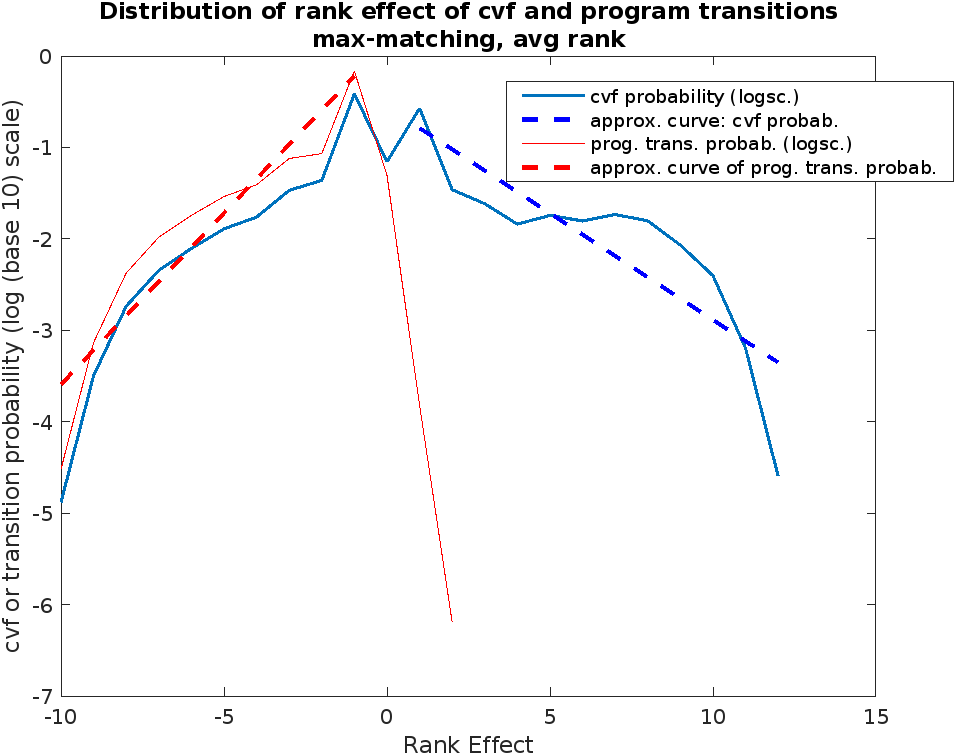}
        \label{fig:histogram-line-max-match-avg}
    }
    \vspace{-10pt}
    \caption{Distribution of program transitions and \cvf{s} by their rank effect. 
    }
    \label{fig:histogram-all}
\end{figure*}

In this section, we evaluate the benefit and the cost of program and \cvf transitions for the three case studies. 
Figure \ref{fig:histogram-all} shows the probability distribution of the rank effect for program and \cvf transitions using \mrank and \arank.


As an illustration of these results in Figure \ref{fig:histogram-line-dijkstra-avg}, we consider the token ring program with 9 processes. We find that 9\% of the \cvf transitions have no effect on the rank. 27\% of \cvf transitions change the rank by at most 1, i.e., they undo the effect of at most one program transition. 
50\% of the \cvf transitions change the rank by at most 5. However, 0.6\% of \cvf{s} change the rank by 50 or more, i.e., to undo their effect, 50 or more program transitions are needed.

Yet another observation from Figure \ref{fig:histogram-all} is that with \mrank, the rank effect of program transitions is always negative whereas with \arank, the effect of program transitions is occasionally positive. (This is expected since \arank may not decrease with every program transition.) 

From Figure \ref{fig:histogram-all},  we find that the distribution of \cvf{s} is exponential in nature in that the fraction of \cvf{s} with cost $c$ decreases exponentially with $c$. 
The approximation curves (straight lines) are obtained in these figures as follows. We logscaled the fraction (probability) of the \cvf{s} observed against their rank effect. The continuous blue and red curves are the curves of \cvf count and program transition count respectively. The dotted blue and red curves are the approximation curves of blue and red continuous curves respectively, modelled as a straight line. 
The approximation for \cvf{s} is done only for the positive domain, that is, only where the positive rank effect is observed. The proximity between the continuous curves and the respective observed curves shows that there is in fact an exponential decrease in the number (fraction) of \cvf{s} with increase in rank effect.


\begin{table}[ht]
    \vspace{-10pt}
    \centering
    \small
    \begin{tabular}{|l|l|}
        \hline
        \textbf{\#nodes} & \textbf{Avg rank}\\
        \hline
        5 & $2.0149 \times~0.10355^{-c}$\\
        6 & $2.5917 \times~0.093247^{-c}$\\
        7 & $1.7705 \times~0.13512^{-c}$\\
        8 & $3.2616 \times~0.093815^{-c}$\\
        9 & $3.4676 \times~0.095589^{-c}$\\
        \hline
    \end{tabular}
    \caption{Coloring: approximation curves - distribution of \cvf{s} by their cost on rank.}
    \label{tab:coloring}
    \vspace{-10pt}
\end{table}

\begin{table}[ht]
    \vspace{-10pt}
    \centering
    \small
    \begin{tabular}{|l|l|}
        \hline
        \textbf{\#nodes} & \textbf{Avg rank}\\
        \hline
        5 & $0.175 \times~0.76807^{-c}$\\
        6 & $0.1516 \times~0.78682^{-c}$\\
        7 & $0.55501 \times~0.63925^{-c}$\\
        \hline
    \end{tabular}
    \caption{Maximal Matching: approximation curves - distribution of \cvf{s} by their cost on rank.}
    \label{tab:max-match}
    \vspace{-10pt}
\end{table}

\begin{table}[ht]
    \vspace{-10pt}
    \centering
    \small
    \begin{tabular}{|l|l|}
        \hline
        \textbf{\#nodes} & \textbf{Avg rank}\\
        \hline
        4 & $0.044826 \times~1.0014^{-c}$\\
        5 & $0.073633 \times~0.80868^{-c}$\\
        6 & $0.056687 \times~0.85615^{-c}$\\
        7 & $0.047871 \times~0.88599^{-c}$\\
        8 & $0.040627 \times~0.90492^{-c}$\\
        9 & $0.035021 \times~0.91867^{-c}$\\
        10 & $0.03495 \times~0.92382^{-c}$\\
        11 & $0.038182 \times~0.92564^{-c}$\\
        12 & $0.038776 \times~0.92929^{-c}$\\
        13 & $0.042963 \times~0.93068^{-c}$\\
        14 & $0.044728 \times~0.93332^{-c}$\\
        \hline
    \end{tabular}
    \caption{Token ring: approximation curves - distribution of \cvf{s} by their cost on rank.}
    \label{tab:token-ring}
    \vspace{-15pt}
\end{table}

We consider a similar analysis for other sizes of the input programs as well. For the token ring program, we identify the distribution of the program and \cvf transitions in Table \ref{tab:token-ring}. From this figure, we observe that the slope of the approximation curve strictly decreases. However, we are not able to observe the same for other programs (coloring (Table \ref{tab:coloring}) and maximal matching (Table \ref{tab:max-match})).

\section{Analyzing Relative Cost of \cvf{s} with respect to Program Transitions}
\label{sec:analyzerelativecost}

As discussed in Section \ref{sec:computecostbenefit}, each program state has a rank (either $\mrank$ or $\arank$). The rank effect of a state change $(s_0,s_1)$ (which could be a program transition or a \cvf) is the difference $rank(s_1) - rank(s_0)$.

While Section \ref{sec:computecostbenefit} focuses on the distribution of individual \cvf and program transitions, in this section, we consider the average effect of program and \cvf transitions. To this end, in Section \ref{sec:relcvf}, we define the relative effect between program and \cvf transitions. 
We describe full and partial analysis in Section \ref{sec:full-partial-analysis-desc.}. We analyze the case studies under these two analysis paradigms. In Section \ref{sec:computerelcvf}, we present the observations that we obtained from the experiments. We analyze these observations and derive approximation curves of \cvf{s} of positive 
rank effect based on the observations.

\subsection{Defining $rel_{cvf}$}
\label{sec:relcvf}

Observe that if $(s_0,s_1)$ is a program transition and $s_0 \in  inv$ then $s_1 \in inv$ and the rank effect is 0. 
Hence for a given program, we evaluate the average of the rank effects of all program transitions that originate outside the invariant (i.e. $s_0 \notin  inv$).
Note that this value, denoted as $\transeff$ is negative since program transitions reduce the rank overall. 
Similarly, we compute $\cvfeff$, the average rank effects of \cvf{s} that begin outside the invariant. As mentioned in Section \ref{sec:cvf}, we only focus on \cvf transitions that increase the rank thereby making the analysis conservative in nature. Therefore, $\cvfeff > 0$.
The relative recovery cost of \cvf{s} is defined as 
$$\relcvf = -\frac{\cvfeff}{\transeff}$$

The relative recovery cost $\relcvf$ measures how many program transitions, on average, are needed to equalize the perturbation effect of a \cvf. 

We evaluate the cost of \cvf{s} and the benefits of program transitions on the same three self-stabilizing programs.
For graph coloring and maximal matching, we used three types of input graph (topology) in our evaluation: ring, power-law graph, and random regular graph.
In a power-law graph, node degrees (number of process' neighbors) follow the power-law distribution and nodes form clusters within the graph. In a random regular graph, nodes have the same degree and are randomly connected. Ring is the special case of random regular graph where every node has degree 2.
We used the tool \textit{networkx} \cite{networkx.duong} to generate those graphs.

\subsection{Full and Partial Analysis} \label{sec:full-partial-analysis-desc.}
For a given case study program, we construct its entire state space and then for each state, we compute its \arank and \mrank. 
For each program transition $(s_0,s_1)$ where $s_0 \notin inv$, we compute its rank effect. 
Then, \transeff is the average of those rank effects. Similarly, we obtain \cvfeff as the average of the rank effects of all  $(s_0,s_1) \in \cvf$ and $s_0 \notin inv$ and $rank(s_1) > rank(s_0)$ . As anticipated, this approach of full analysis suffers from the state space explosion problem. For example, full analysis of the token ring program with 15 processes easily exhausts 170 GB of memory. Hence, we also consider partial analysis where we compute the ranks of a number of random states without constructing the whole state space. Suppose $s_0$ is a randomly chosen state, we probe a pre-determined number of random paths from $s_0$ to the program's invariant. The maximum and average length of these selected paths is the approximation of $\mrank(s_0)$ and $\arank(s_0)$, respectively. Subsequently, given the rank of two states $s_0$ and $s_1$ obtained in this way, the rank effect of a program transition (or \cvf) $(s_0,s_1)$ is the difference of the ranks of $s_1$ and $s_0$.



\subsection{Computing $rel_{cvf}$}
\label{sec:computerelcvf}

\begin{figure}
    \vspace{-5pt}
    \centering
    \includegraphics[width=0.4\textwidth]{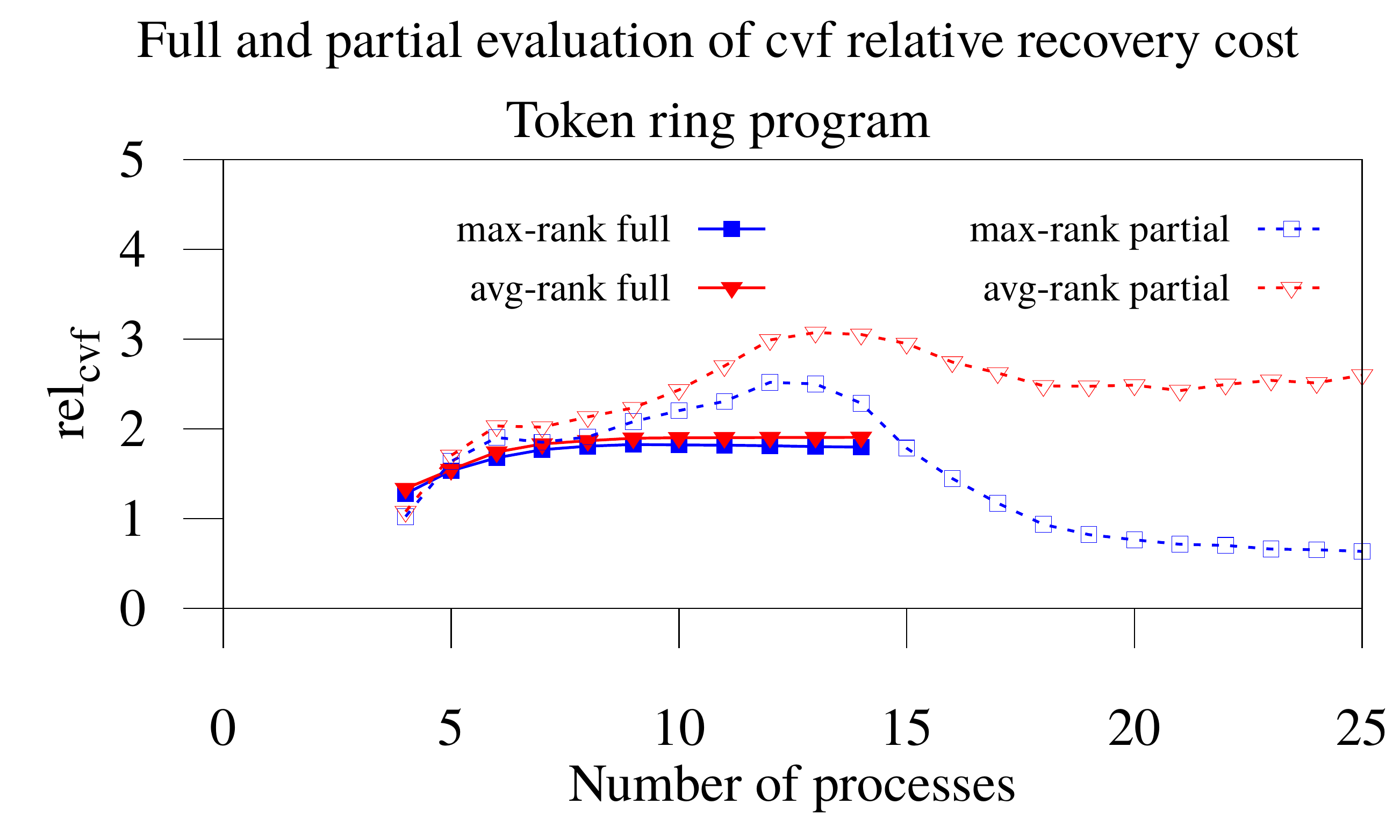}
    \label{fig:dijkstra}
    \vspace{-5pt}
    \caption{The relative recovery cost of \cvf{s} in the token ring program.}
    \label{fig:cvf-recovery-cost-dijkstra}
    \vspace{-10pt}
\end{figure}

\begin{figure*}[ht]
    \centering
    \subfloat[]{
        \includegraphics[width=0.33\textwidth]{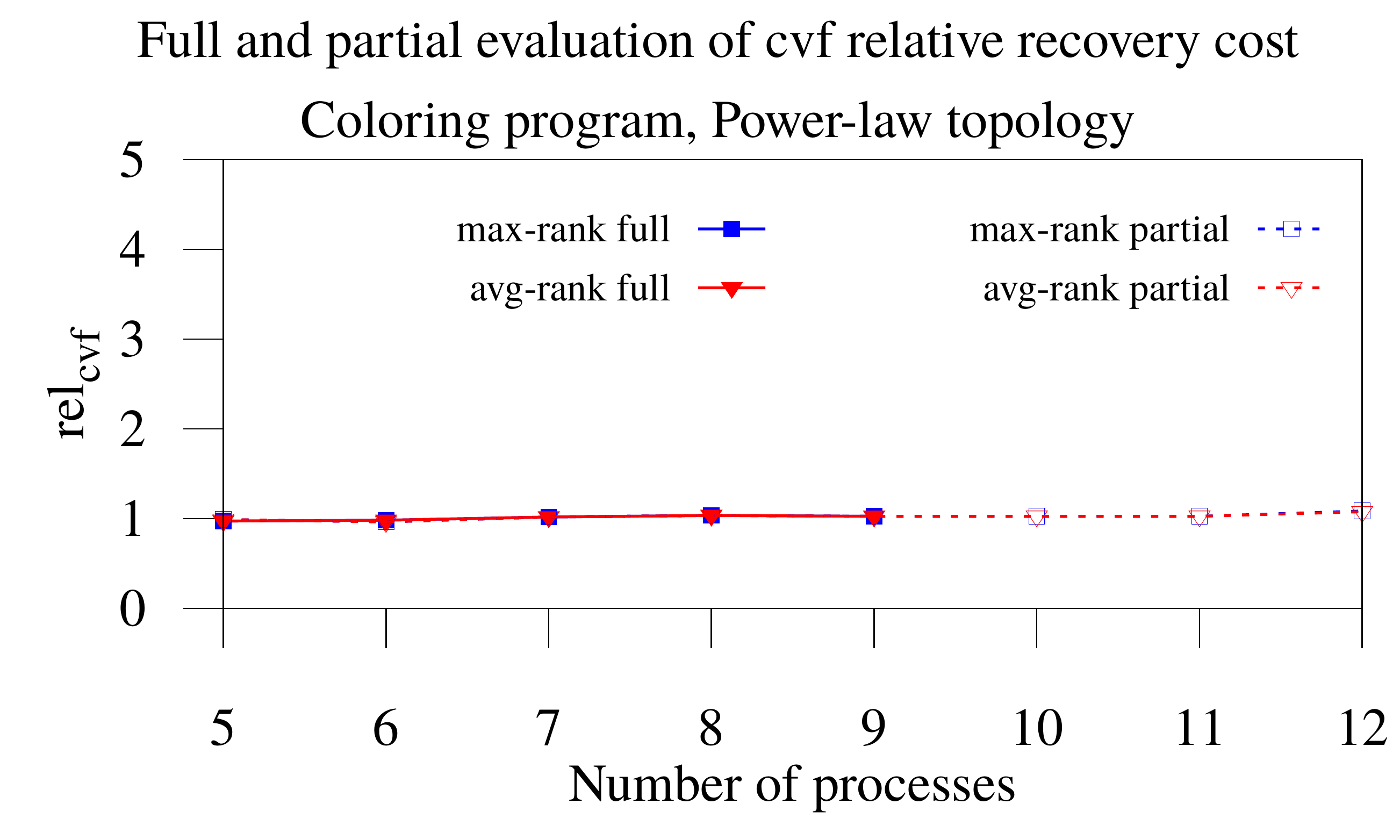}
        \label{fig:color-powerlaw}
    }
    \subfloat[]{
        \includegraphics[width=0.33\textwidth]{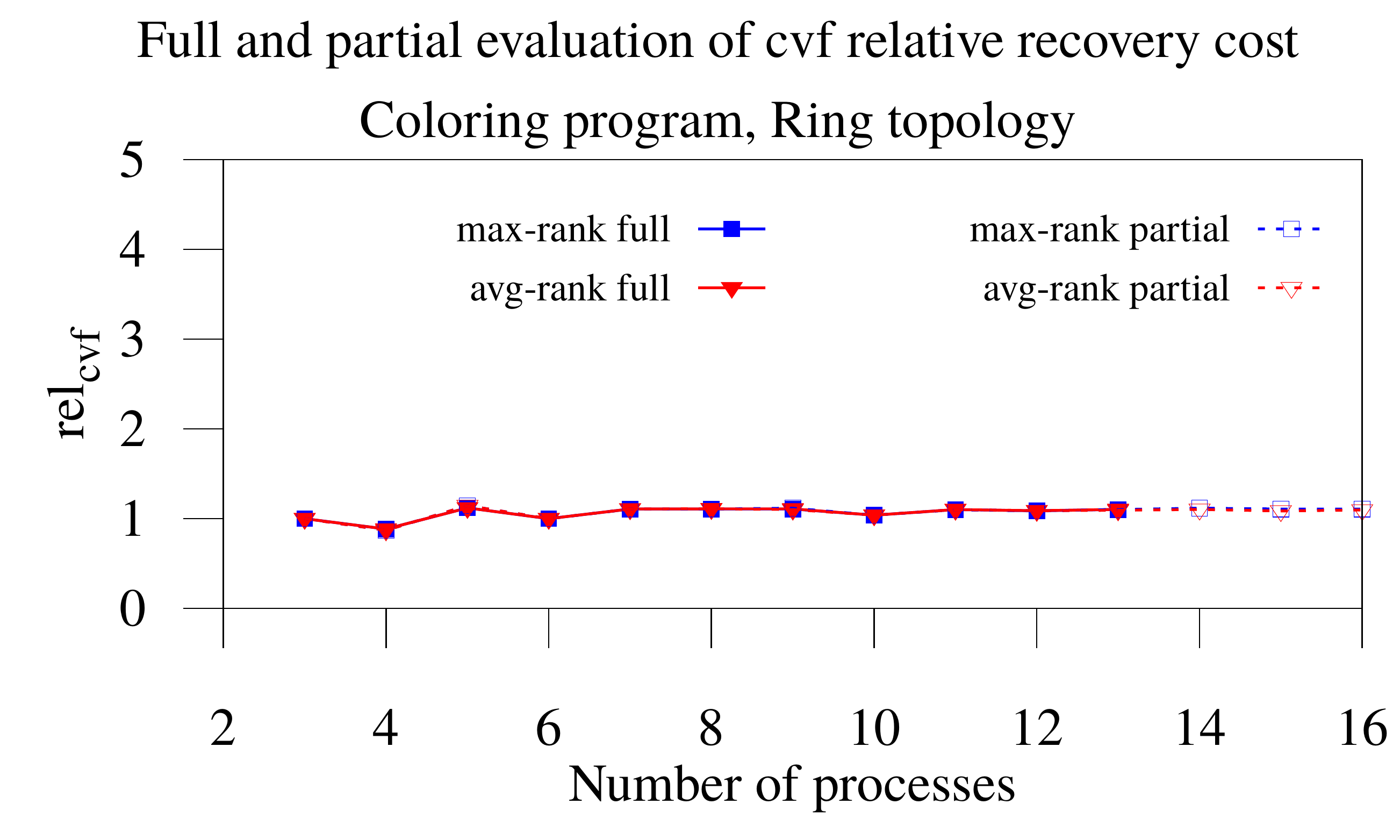}
        \label{fig:color-ring}
    }
    \subfloat[]{
        \includegraphics[width=0.33\textwidth]{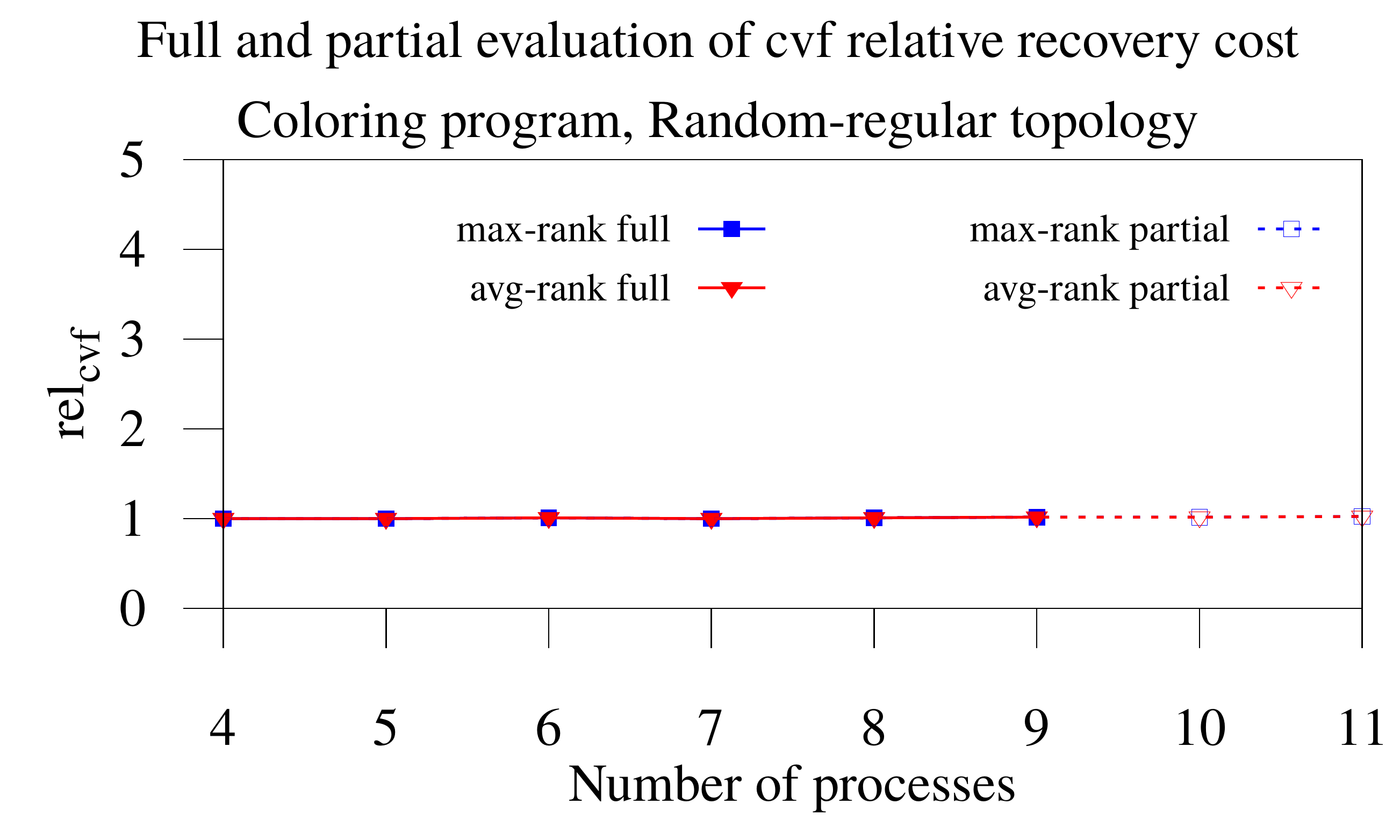}
        \label{fig:color-rand-reg}
    }
    \\
    \vspace{-5pt}
    \subfloat[]{
        \includegraphics[width=0.33\textwidth]{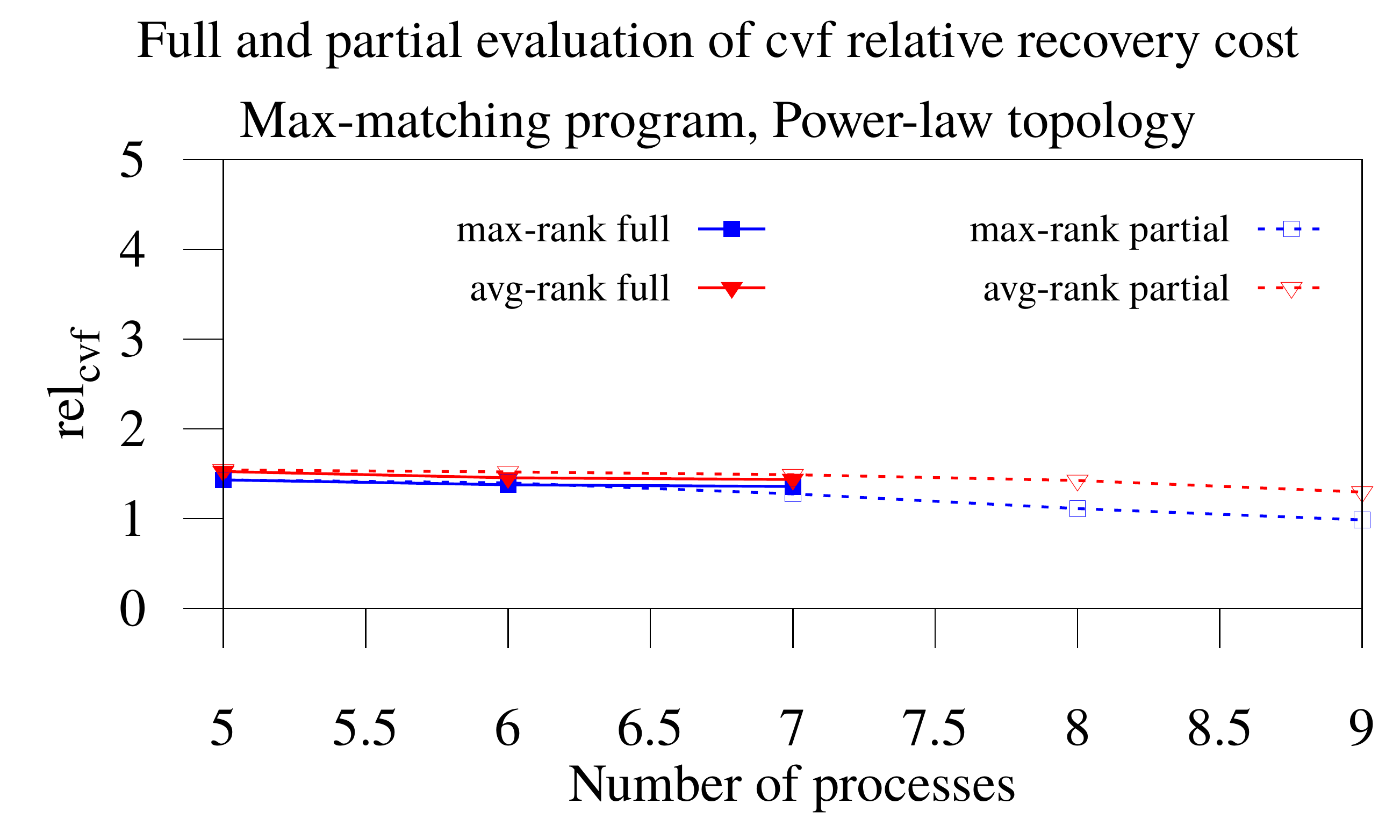}
        \label{fig:max-matching-powerlaw}
    }
    \subfloat[]{
        \includegraphics[width=0.33\textwidth]{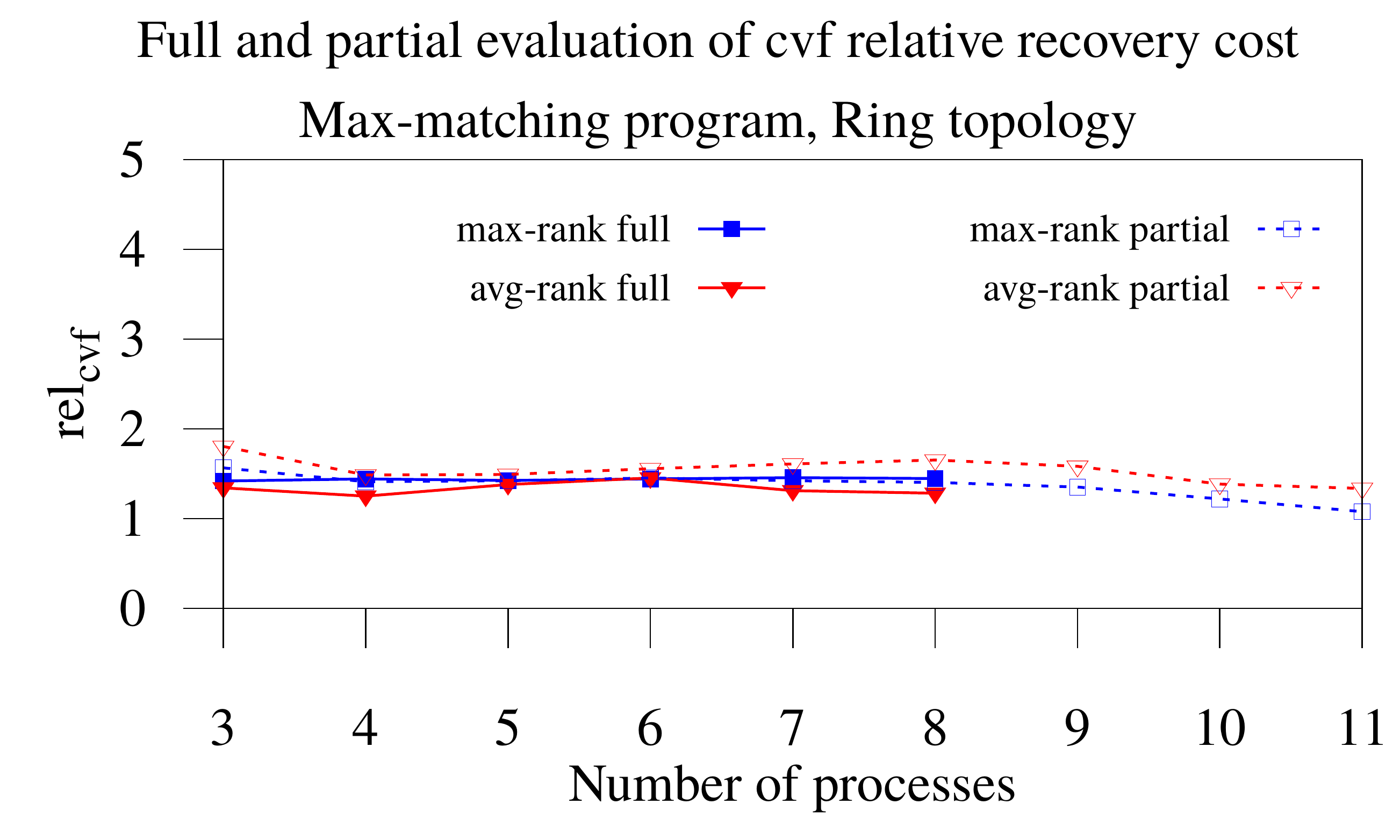}
        \label{fig:max-matching-ring}
    }
    \subfloat[]{
        \includegraphics[width=0.33\textwidth]{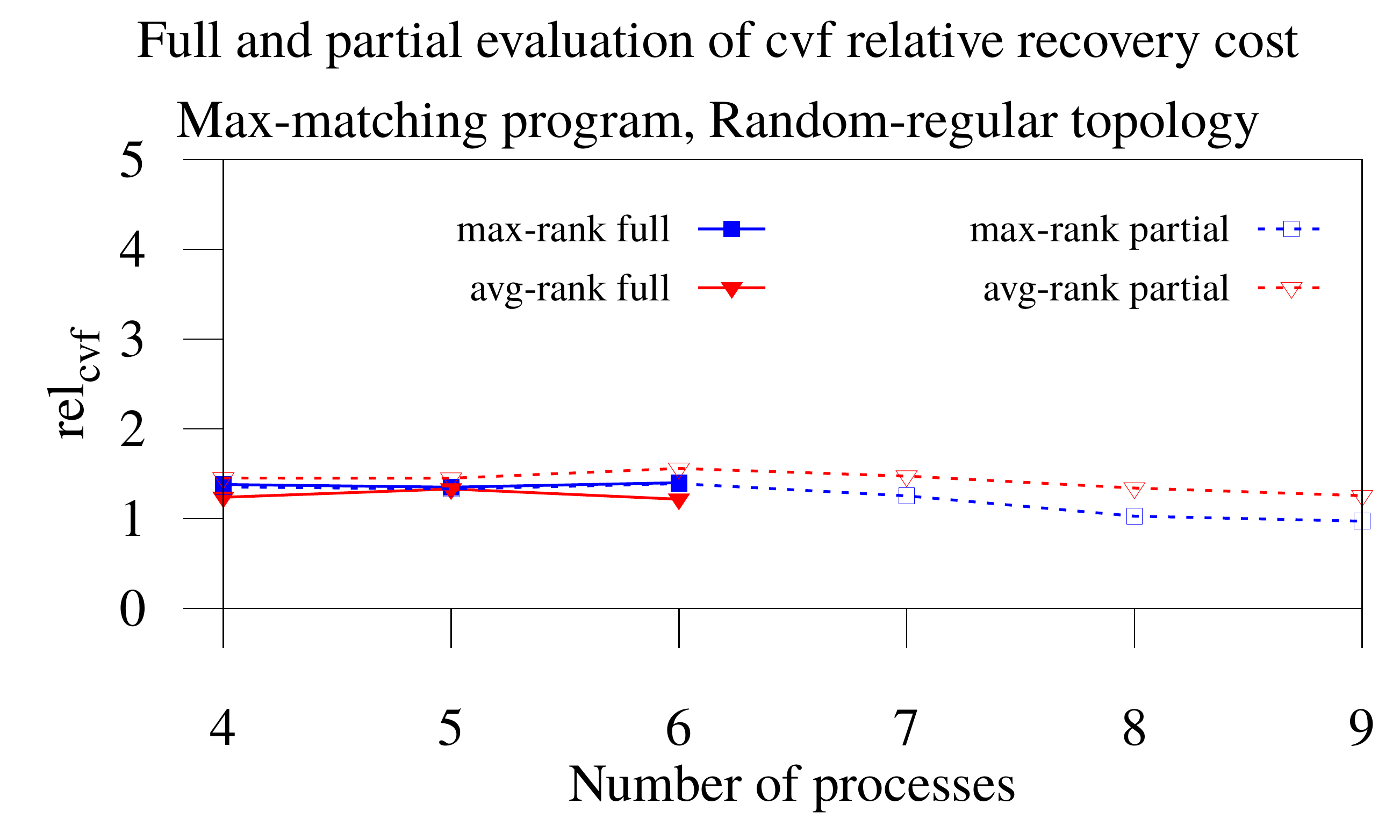}
        \label{fig:max-matching-rand-reg}
    }
    \vspace{-5pt}
    \caption{The relative recovery cost of \cvf{s} in coloring and maximal matching programs.}
    \label{fig:cvf-recovery-cost-all}
    \vspace{-5pt}
\end{figure*}

\textbf{Analysis with \mrank and \arank. }From Figures \ref{fig:cvf-recovery-cost-dijkstra} and \ref{fig:cvf-recovery-cost-all},
we observe that the \cvf relative recovery cost $\relcvf$ does not significantly differ between $\mrank$ and $\arank$ in full analysis. 
In other words, when we utilize full analysis, the number of program transitions required to neutralize a \cvf remains the same whether we use \mrank or \arank.

\textbf{Partial Analysis vs Full Analysis. } 
Figures \ref{fig:cvf-recovery-cost-dijkstra} and \ref{fig:cvf-recovery-cost-all} also consider the analysis via partial analysis and full analysis. When full analysis is feasible, we conducted both full and partial analysis. Otherwise, we only considered partial analysis.
When both are possible, they are consistent with each other. This indicates that partial analysis provides reasonably good estimates of full analysis. We also find that partial analysis via \arank is more stable and follows the trend of full analysis better than the analysis of $\mrank$.
This is expected because the sampling method generally provides a good estimate of the average value but not the extremum. This is beneficial since the average-case cost of \cvf{s} is more relevant regarding the convergence to legitimate states.
The results also imply that we can get a good estimate of \relcvf when the state space is large.

\textbf{Observations about values of $\relcvf$. }
Although the specific pattern and values of $\relcvf$  depend on given case study, we observe that the value of $\relcvf$ is small (roughly 2-4) indicating that the stabilization property is likely to be (probabilistically) preserved even if \cvf{s} are frequent and only a small number (2-4) of program transitions execute between \cvf{s}. 
Also, the value of $\relcvf$  does not vary significantly with the number of processes indicating that analysis of a problem with a small number of processes could generalize to a large number of processes.






\section{Performance Analysis in the Presence of \cvf{s}}
\label{sec:performance}




In this 
section, we validate the observations about $\relcvf$ to determine the ability of a program to converge in the presence of \cvf{s}. If $\relcvf=3$, this indicates that a \cvf undoes the progress achieved via 3 program transitions. Hence, if more than 3 program transitions are executed between \cvf{s}, it is anticipated that the recovery to legitimate states would still occur. Of course, if the number of program transitions between \cvf{s} is just above 3 then there may be a substantial increase in the number of steps required for convergence. We evaluate this hypothesis with simulations of the given case study programs in the presence of \cvf{s}.

We choose simulations to validate the hypothesis because simulations allow us to control the rate of \cvf{s}. By contrast, in an actual experiment on real networked systems, we cannot easily change the rate of \cvf{s} 
since we cannot control when two neighboring processes access the same data item simultaneously
.  Furthermore, in an actual system, methods used to detect \cvf{s} interfere with the actual computation.
In simulations, we introduce the parameter $\cvfintv$ to denote the average number of program transitions between \cvf{s}. For example, $\cvfintv = 4$ means (on average) 4 program transitions execute between consecutive \cvf{s}.

We conduct our simulations as follows: We randomly select an initial state, say $s_0$, and 
emulate 
a computation of the given program from that initial state.  
In order to emulate
such a computation, in each iteration, we execute a random enabled action. 
Furthermore, in each iteration, we randomly choose a process and perturb the state of that process in such a way that mimics the effect of a \cvf. The parameter $\cvfintv$ determines how frequently \cvf{s} are introduced. We terminate the execution when the program state is legitimate (inside the invariant) or when the number of iterations exceeds a threshold.
In the latter case, we treat it as if the program has failed to converge in the presence of \cvf{s}. 
Let $\convstep$ denote the number of iterations by the time of termination.
For each initial state ($s_0$), we consider five computations and take the average of $\convstep$. 
Subsequently, for the same initial state, we compute $\convstep$ for different values of $\cvfintv$.

\begin{figure*}[!t]
    \centering
    \subfloat[Token ring]{
        \includegraphics[width=0.31\textwidth]{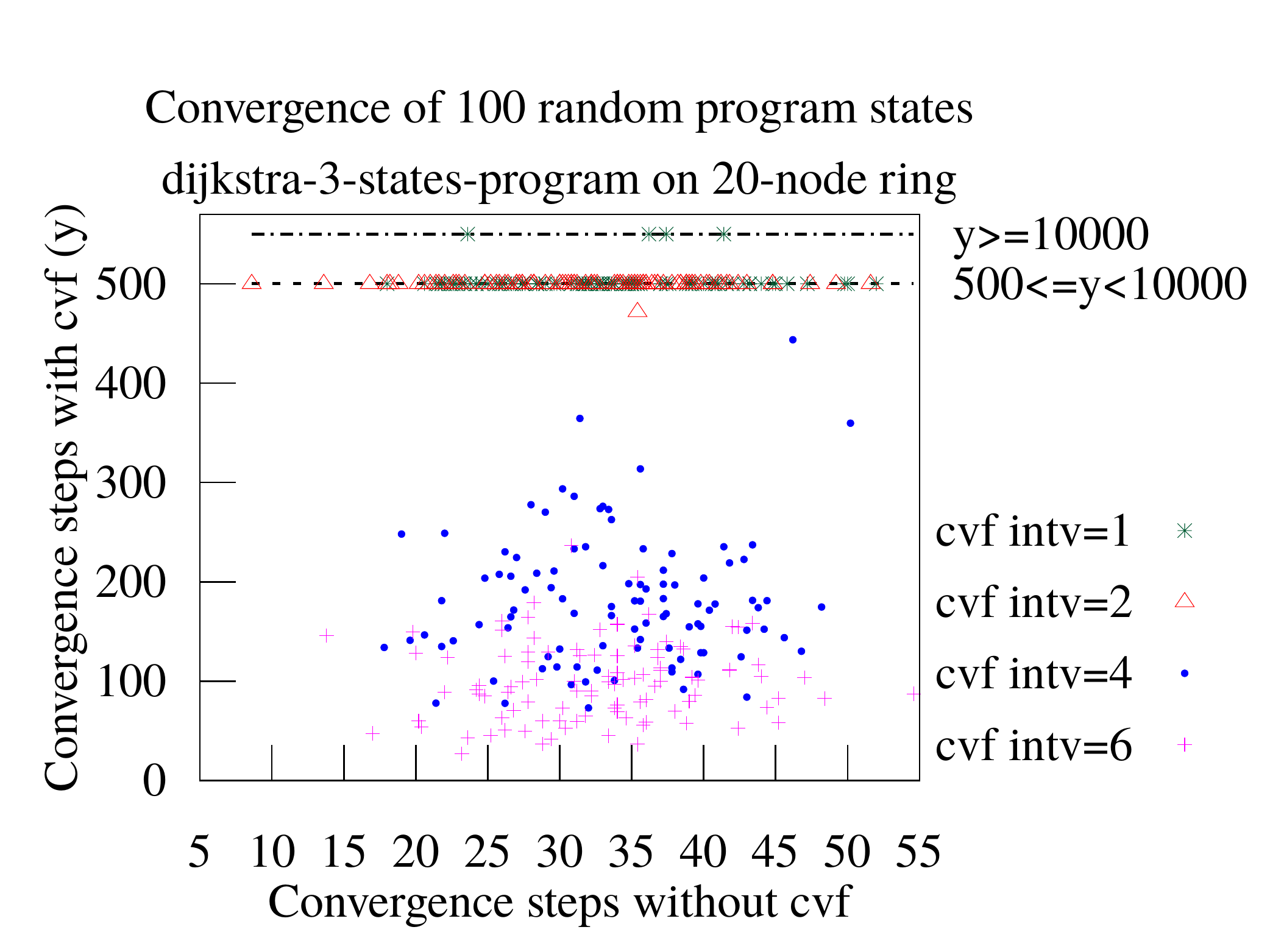}
        \label{fig:detailed-conv--dijkstra}
    }
    \subfloat[Coloring]{
        \includegraphics[width=0.31\textwidth]{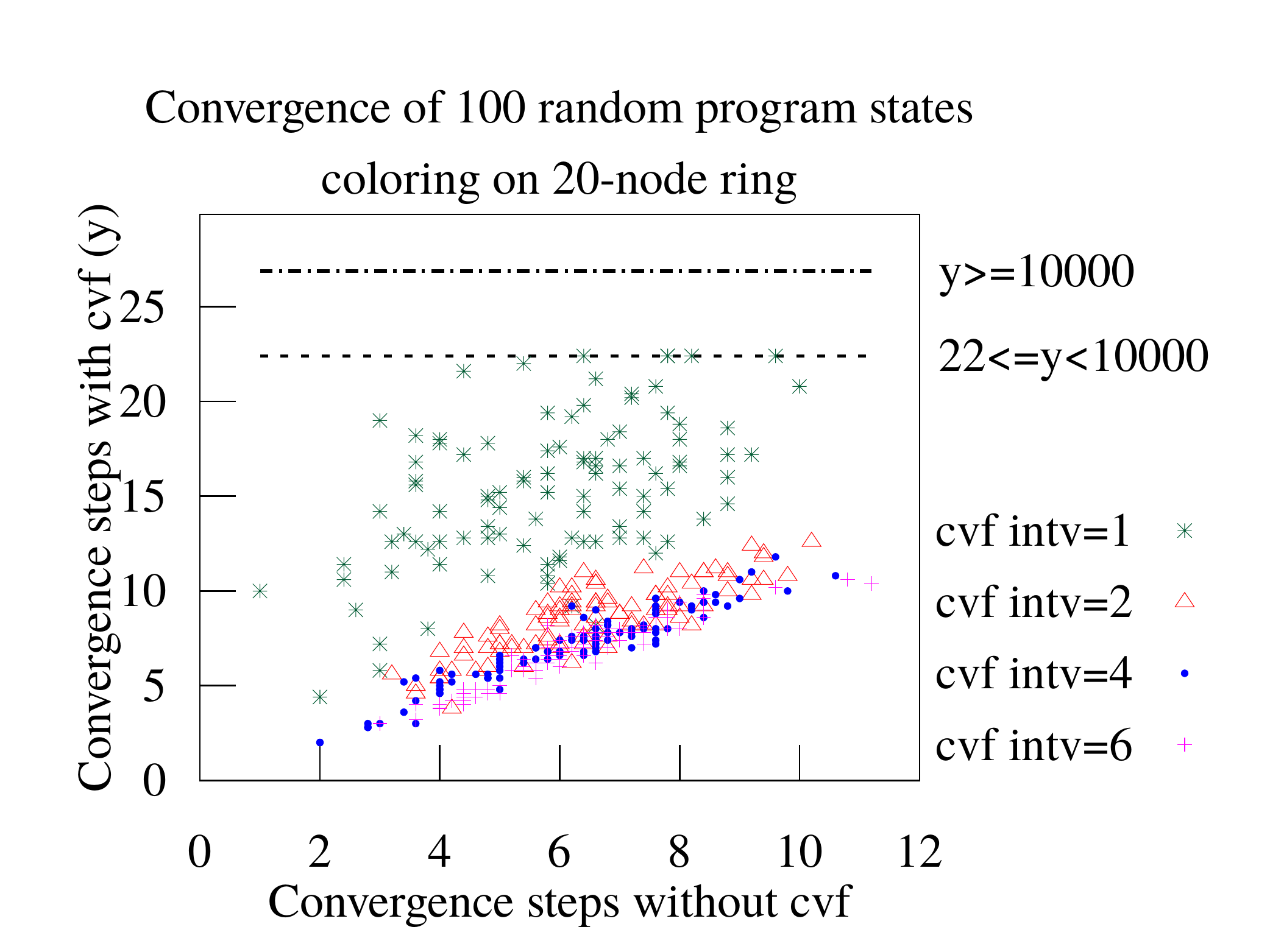}
        \label{fig:detailed-conv--coloring}
    }
    \subfloat[Max-matching]{
        \includegraphics[width=0.31\textwidth]{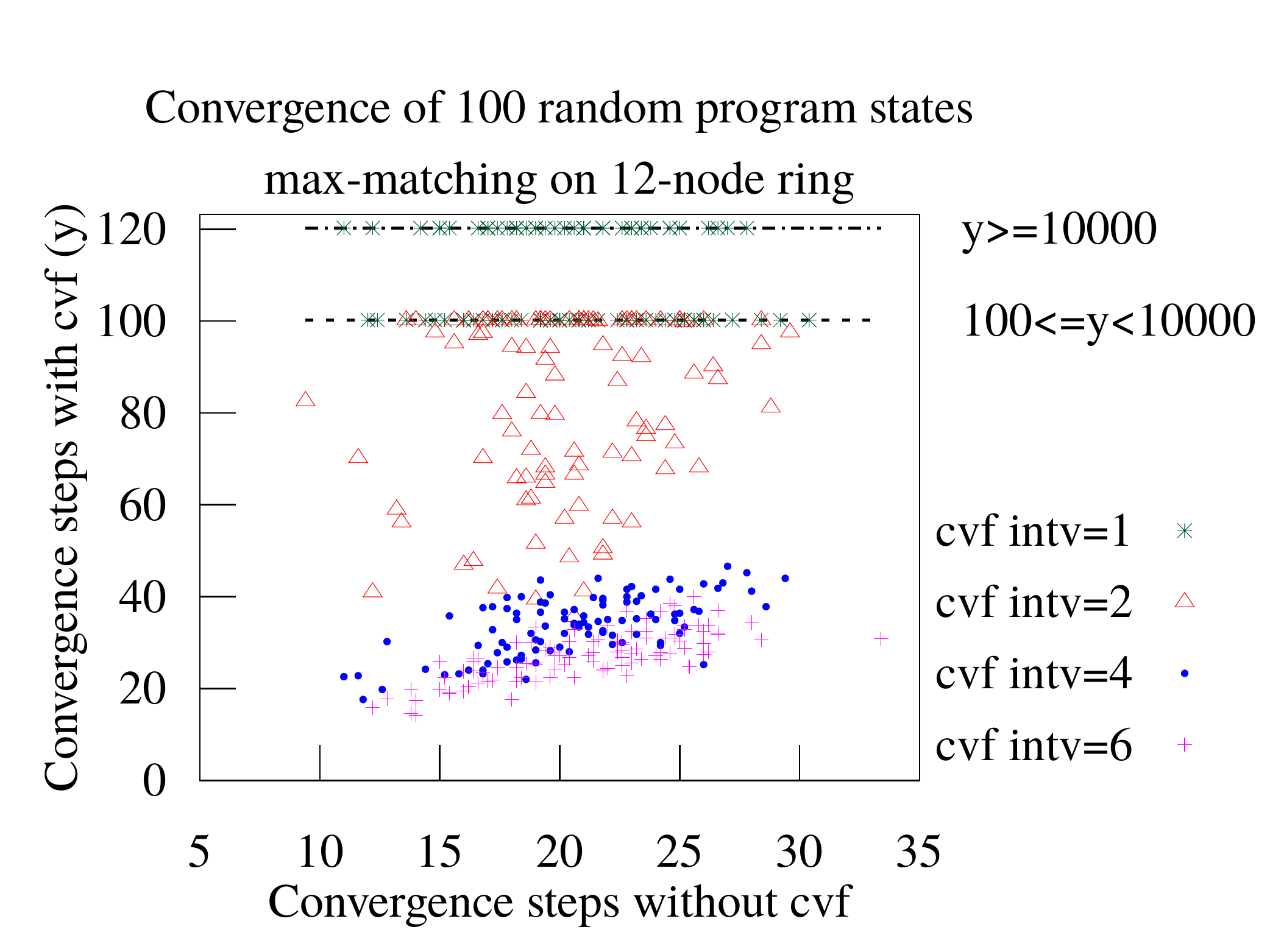}
        \label{fig:detailed-conv--max-matching}
    }
    \caption{The convergence of three case study programs on ring graphs. Each point is a random initial state outside the invariant. The x-coordinate (y-coordinate, respectively) is the average number of transitions needed for the program to converge from that state to the invariant when there is no \cvf (when there are \cvf{s}, respectively). A value of 10,000 on y-coordinate means the program does not convergence in the allocated time.}
    \label{fig:detailed-conv}
\end{figure*}

\begin{figure*}[!t]
    \centering
    \subfloat[]{
        \includegraphics[width=0.31\textwidth]{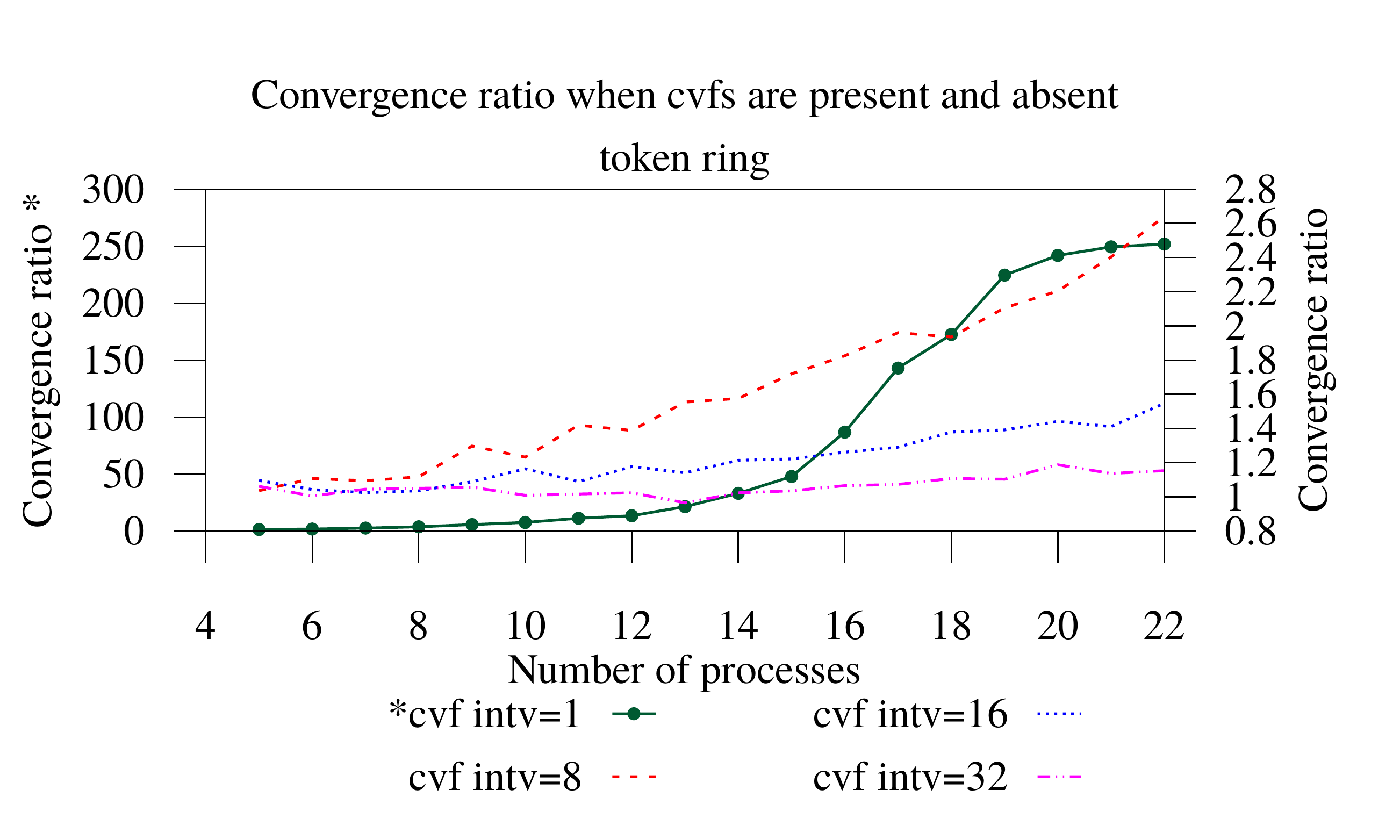}
        \label{fig:simulation-dijkstra}}
    \subfloat[]{
        \includegraphics[width=0.31\textwidth]{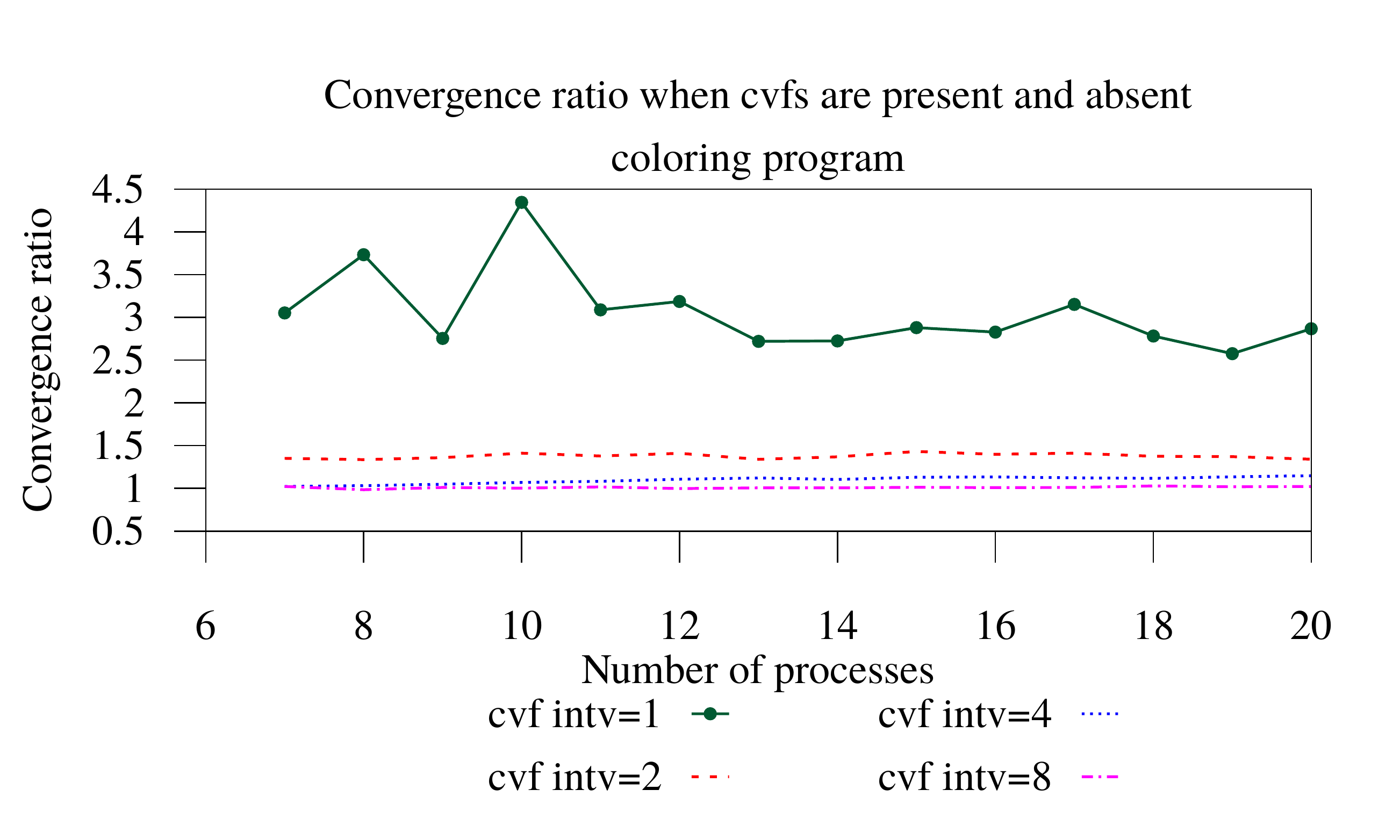}
        \label{fig:simulation-color-ring}
    }
    \subfloat[]{
        \includegraphics[width=0.31\textwidth]{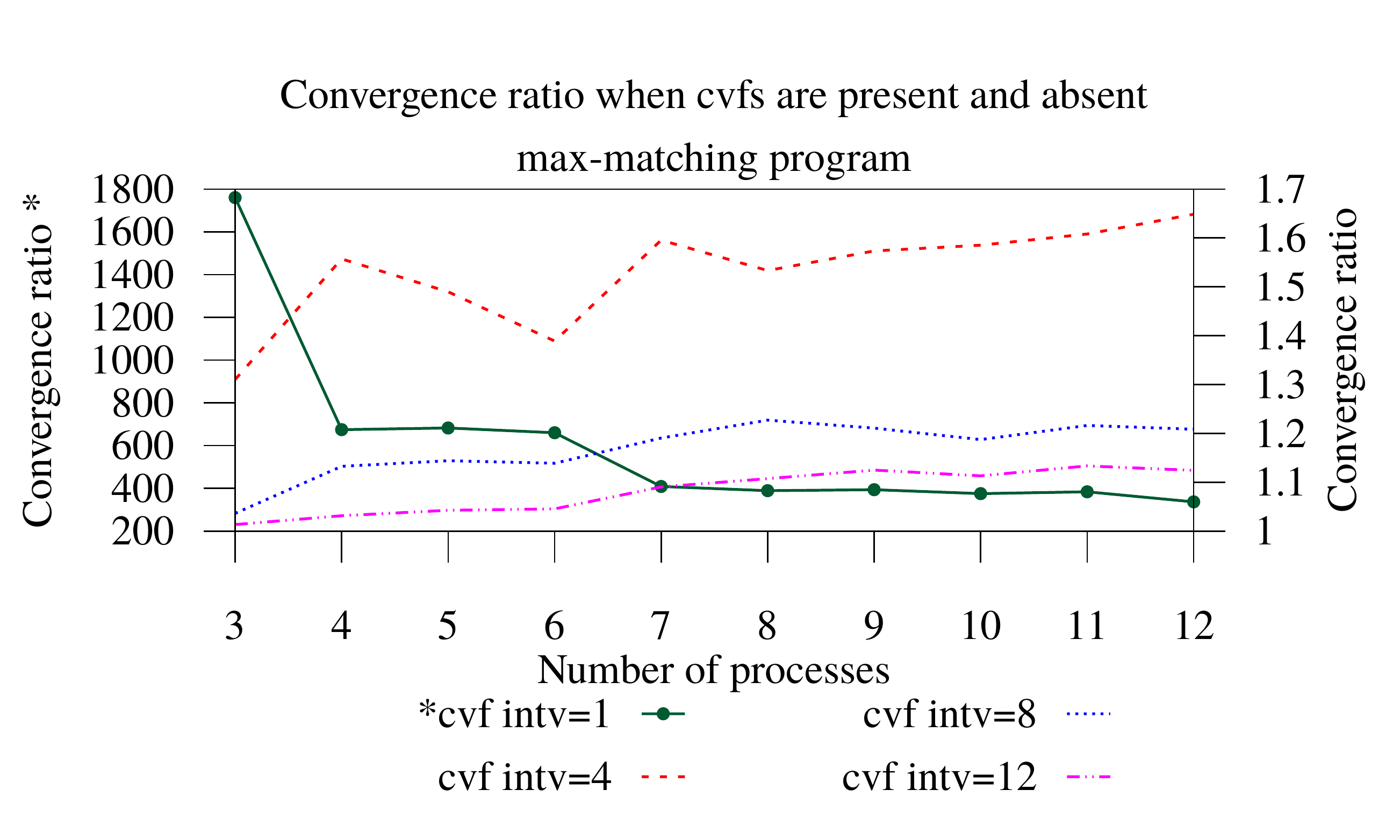}
        \label{fig:simulation-max-match-ring}
    }
    \caption{Simulations: The ratio between the number of convergence steps of three case study programs when \cvf{s} are present and absent.}
    \label{fig:simulations}
\end{figure*}

We present the data from our simulations in Figure  \ref{fig:detailed-conv} as a scatter plot and in Figure \ref{fig:simulations} as the relative increase in the number of steps for convergence due to \cvf{s}. 
In Figure \ref{fig:detailed-conv}, each point is the simulation results of a given program for an initial state with a specific \cvfintv value. A point with coordinate $(X, Y)$ indicates that for the initial state considered, it took on average $X$ steps ($Y$ steps, respectively) for the program to converge in the absence (in the presence, respectively) of \cvf{s} with the given value of \cvfintv.
To present the data in an easy-to-read fashion, we remove certain details when $Y$ coordinate is large. For example, in Figure \ref{fig:detailed-conv--dijkstra}, we group the points when $Y$ coordinate is between 500 -- 10000 (threshold) to indicate that convergence was achieved but it took too long. 


Figure \ref{fig:simulations} presents the simulation data to compare the number of steps for convergence in the absence of \cvf{s} and in the presence of \cvf{s}. 
If $\cvfintv$ is small then the number of steps to converge could be very large (or infinite). This is observed in Figure \ref{fig:simulation-dijkstra} where the number of steps increases by a factor of $100$ or more if $\cvfintv=1$.
However, if $\cvfintv=8$, the number of steps for convergence in the presence of \cvf{s} is about $1.2-2.6$ times more than the number of steps in the absence of \cvf{s}. This is consistent with the analysis in Figure \ref{fig:cvf-recovery-cost-dijkstra} for the token ring program where the ratio $\relcvf$ is about 3. 
That means that a \cvf negates the benefit of (approximately) 3 program transitions. Hence, the benefit of 8 program transitions and 1 \cvf is (approximately) the same as 5 program transitions, thereby increasing the number of steps to be $\frac{8+1}{5}$ (i.e., $1.8$) times. 

Figure \ref{fig:simulation-color-ring} analyzes the coloring program. Here, from Figure \ref{fig:cvf-recovery-cost-all}, the program is expected to converge even if a \cvf occurs between two consecutive program transitions. Figure \ref{fig:simulation-color-ring} confirms this anticipation and finds that the increase in the number of steps is approximately 3 times if $\cvfintv=1$.
If \cvf frequency is reduced then the increase in the number of steps drops quickly. For example, if $\cvfintv=8$ then there is virtually no increase in the number of steps for convergence. 
Figure \ref{fig:simulation-max-match-ring} considers the same question for the maximum match problem. As expected, when \cvf{s} are too frequent ($\cvfintv=1$), the increased steps is very high. But when \cvf frequency is reduced ($\cvfintv\geq 8$), the increase is only $0-20\%$. 

Note that it is anticipated that $\convstep$ will increase in the presence of \cvf{s}. However, each step is expected to be faster, as processes do not need to synchronize with each other. 

\section{Interpreting the Results of Case Studies and Their Implications }
\label{sec:interpret}

As discussed in the Introduction, existing approaches for 
concurrent computing focus on ensuring that the program state is always consistent. 
In other words, these approaches rely on preventing \cvf{s}. However, to use these approaches in practice, it is necessary for coordination among threads in the given program. By contrast, permitting and tolerating \cvf{s} allows one to eliminate synchronization between threads as each thread could work independently. 

\subsection{Number of Steps vs Time}
\label{sec:steps}

In this work, we intentionally focused on number of steps required for convergence instead of time.  Note that with \cvf-tolerance, we can run the program without the necessary coordination/synchronization.  Hence, we expect that 
if the program is run with $n$ concurrent threads, we can get a speedup of $n$ (subject to suitable load balancing). By contrast, without \cvf-tolerance, we will need to ensure that execution of each action is atomic and therefore will suffer from large overhead of lock/unlock. 

Additionally, when comparing \cvf-prevention and \cvf-tolerance, consider the case where \cvf-prevention is implemented using local mutual exclusion. Specifically. if node $j$ is executing at a given time none of its neighbors can execute at that time. Thus, if we have the underlying graph to be a 2-dimensional grid then at most half of the nodes can execute at a time. This means that as long as the number of steps does not increase 2-fold, \cvf-tolerance is expected to provide a benefit over \cvf-prevention.
Taking this further, one would expect that at most $\alpha$ nodes can execute at a given time where $\alpha$ is the independence number (size of the maximum independent set) of the communication network among the processes. Hence, as long as the number of steps does not increase by a factor of $\frac{n}{\alpha}$, \cvf-tolerance is expected to be better.

When we eliminate \cvf{s}, an action at node $j$ would require that $j$ first lock the relevant data then execute the action and then release the lock. There are two types of overheads in this scheme. First, there is an overhead for obtaining and releasing the locks even if there is no contention. Second, there is the required waiting if the locks are not available when requested. 
To estimate these overheads, we considered the execution of graph coloring program \cite{GT2000OPODIS.short} using only one thread- with and without the necessary locks.
Since our goal was to evaluate the overhead, we ran the program with 1000 steps. Since there is only one thread, the overhead is only for acquiring locks (as there is no contention). 
Even under these circumstances, the time of execution was more than 2-fold when locks were acquired before every step. The actual overhead would be higher when we account for contention and where some threads need to wait on acquiring the locks.

Our results from Section \ref{sec:performance} show that in most cases, the increased number of steps is very small (typically, less than 20\%). Thus, the increased number of steps would not significantly affect the performance when compared with the approach of \cvf-prevention. 

We anticipate that a slight variation between the overhead (e.g., 10\% increased steps vs 20\% increased steps) would not substantially affect the performance. By contrast, in a handful of cases where the rate of \cvf{s} is very high (e.g., token ring program with one \cvf between two program actions), \cvf-prevention is preferable to \cvf-tolerance.

\subsection{Implication for Stabilizing Programs}
Our focus in this paper was in stabilizing programs that, by definition, tolerate \cvf{s}. Specifically, if the number of \cvf{s} is finite, the program is (deterministically) guaranteed to converge. And, if the frequency of \cvf{s} is small, the program is guaranteed to converge probabilistically. 

When we view tolerating \cvf{s} through the lens of algorithms such as matching and coloring, we note that these algorithms are 
silently stabilizing
. In other words, once the program reaches the state where valid matching or coloring is found, there is no need to execute any actions further. Such programs benefit the most from \cvf-tolerance, as they can continue to run until they reach a legitimate state. At this point, \cvf{s} (which are caused by reading old values of variables) have no effect, as the program state remains unchanged forever. 

Programs such as token rings are not terminating, i.e., they continue to execute even after they reach legitimate states. If \cvf{s} occur when the program is in a legitimate state, it is possible that the \cvf{s} may cause the resulting state to be illegitimate. If this happens, the program will still return to legitimate states. However, this behavior is different from the behavior of stabilizing token ring program which guarantees that once a legitimate state is reached, subsequent computation remains in legitimate states in the absence of faults. 
In other words, once legitimate states are reached, we should prevent \cvf{s} rather than tolerate them. However, outside the legitimate states the program should permit and tolerate \cvf{s}. Thus, the execution can be partitioned into two phases (1) quiescent phase where the algorithm has recovered to legitimate state(s) and no recovery actions are needed, (2) recovery phase, which is initiated by some node due to an observed inconsistency. 



For any self-stabilizing algorithm, it is desirable that the time spent in the recovery phase is as minimal as possible. The results in this paper allow us to reduce the time spent in the recovery phase. The intuition of how to achieve this is as follows:

If a node enters the recovery phase, it should transition itself to quiescent phase after a certain time, say $\tau_1$. When a node enters the quiescent phase, it should remain there for a certain duration, $\tau_2$, to prevent constant transition from recovery to quiescent phase (and vice versa). 
Furthermore, in recovery phase, the execution  should permit and tolerate \cvf{s}. By contrast, in quiescent phase, the execution should be \textit{conservative}, i.e., it should eliminate \cvf{s}.  (Note that \cvf{s} are possible even if a single node is in the recovery phase.)

Observe that as long as $\tau_2$
is large enough, recovery to legitimate states is guaranteed.
Furthermore, as long as the program is in a legitimate state and each process is in the quiescent phase, the program would remain in a legitimate state. Additionally, as long as \cvf{s} are rare, during the recovery phase, faster execution to the invariant would be possible. From the analysis, we observe that the actual perturbation caused by a \cvf is typically very small, just a handful of program transitions are sufficient to neutralize its effect. 
In turn, this implies that the time spent by the program outside legitimate states can be reduced if the program permits \cvf{s} when the state is \textit{perceived} to be outside legitimate states and eliminates \cvf{s} when the state is \textit{perceived} to be inside legitimate states.

\section{Related Work}
\label{sec:related}

In this work, we focused on permitting and tolerating \cvf{s} rather than preventing them as done in most concurrent algorithms. The closest work to permitting and tolerating \cvf{s} is that by Garg \cite{garg20} 
on Lattice Linear Problems. A key property of these programs is that the set of reachable states form a lattice with one optimal state and it is guaranteed that in any suboptimal state, there is some node whose execution is critical. If this node executes with old information about others, it will either perform the correct action or perform no action. Thus, when this critical node executes, the program will reach closer to the optimal state. Due to the lattice structure, it is guaranteed that the program will reach the optimal state even if any number of \cvf{s} occur. Unfortunately, the requirement of the lattice structure makes it impossible to use it in various problems where underlying state space does not form a lattice.

Stabilizing programs \cite{EDW426}, by contrast, tolerate \cvf{s} if they are either terminating or infrequent. In other words, their tolerance of \cvf{s} is less than that for lattice linear programs. However, they are applicable for a large class of problems. 
Stabilizing algorithms have been designed for spanning trees \cite{ag94} leader election \cite{ADDDL2017IPDPS}, matching \cite{IOT2019PODC}, dominating set \cite{KKM2017SSS}, clustering 
\cite{DBLP:journals/tcs/DattaDHLR16}, etc. 

In \cite{NK2020SRDS.short}, the concept of consistency violation fault was introduced as a formalization of the program perturbations that occur when local mutual exclusion is removed. 
{While the experimental results of \cite{NK2020SRDS.short} demonstrated an improvement in performance, the reasoning behind it was unclear
}
To the best of our knowledge, this paper is the first work that formally analyzes the effect of \cvf{s} during the recovery phase of self-stabilizing programs to show that the effect of \cvf{s} is small and can be computed for a given program. It also shows the rate of \cvf{s} for which overall recovery would be faster. In turn, it can assist designers to determine the best approach for reducing convergence time for a given self-stabilizing program.

\section{Conclusion}
\label{sec:concl}

Execution of concurrent program has typically relied on ensuring that the program state always remains consistent. This requirement is enforced through various mechanisms such as linearizability, local mutual exclusion, etc. Thus, even when concurrent operation is permitted, the goal is to ensure atomicity so that the operations can be ordered in some fashion. 

Our focus is on an alternative approach. Specifically, we focus on permitting and tolerating consistency violation faults (\cvf{s}) rather than eliminating them as it is done in traditional literature. There is empirical evidence that such an approach can significantly enhance the performance, \textit{if the program is designed to tolerate \cvf{s}}. However, the only evidence of this is via experimental analysis that makes it difficult to determine the expected benefit of \cvf-tolerance without experimental evaluation. 

In this paper, we demonstrated that the results of experimental evaluation can be predicted by static analysis of the program state space. Furthermore, even partial analysis provides sufficient details to determine if \cvf-tolerance would be beneficial. Towards this end, we focused on analyzing three stabilizing programs (that naturally tolerate \cvf{s})  to evaluate how \cvf{s} affect them. 



We analyzed the cost of \cvf{s} in the steps required for recovery (i.e., to reach legitimate states from an arbitrary state) in these case studies. To this end, we defined the notion of \arank and \mrank that identify how far a given state is from a legitimate state (e.g., a state with valid coloring). We evaluated the rank change by \cvf{s} and by program transitions. We observed that the cost of \cvf{s} follows an exponential distribution. In other words, the probability that a \cvf increases the rank by $c$ is $A B^{-c}$, where $A$ and $B$ are constants. This indicates that while most \cvf{s} have a minimal cost on recovery, a few \cvf{s} can cause a significant perturbation. We also observed that the benefit of program transition (i.e., rank change by the program transition) also follows an exponential distribution.



We also computed the relative cost of \cvf when compared with program transitions. Specifically, we consider $\relcvf$ which denotes the number of program transitions that would be needed to compensate for a single \cvf. Specifically, this is computed by treating as if all \cvf{s} and program transitions are equally likely and taking the ratio of the average rank change by a \cvf and the average rank change by a program transition. We compute $\relcvf$ by either full analysis of the program state space or via partial analysis. The former is more accurate but suffers from state space explosion. We find that the computation of $\relcvf$ by both analysis match when we can utilize both approaches. This indicates that partial analysis (especially with \arank) would suffice to compute the relative effect of \cvf. 

We also evaluated the actual effect on recovery in the presence of \cvf{s} and compare it to $\relcvf$. As  $\relcvf$ characterizes the number of program transitions needed to compensate the effect of a single \cvf, if $\relcvf$ program transitions occur between \cvf{s} then we expect that program transitions will not make substantial progress. On the other hand, if substantially more program transitions execute between \cvf{s} (e.g., 2 $\relcvf$ program transitions execute between \cvf{s})) then the program continues to converge without a significant increase in the number of steps for convergence.


We have developed a toolset that is generic so that by only coding the specific algorithm, one can obtain the effect of \cvf on that program. Specifically, our analysis of \mrank~and \arank~uses program transitions as a parameter. Thus, simply specifying the new program transitions, one can analyze 
the effect of \cvf{s}
for the given program. We intend to provide this as a toolset for the community to use.

There are several future work in this area. One is to relate the occurrence of \cvf{s} with the level of 
concurrency and the size of the input. A \cvf occurs due to non-atomic execution of program actions.  Thus, it is likely to increase with level of concurrency 
(more conflict when the number of concurrent threads increases but size of the input is kept constant) and decrease with the size of the input (less chance of a conflict if the input size is increases but the available threads are kept constant). Another future work is developing Markov chain model using the exponential nature of the rank change for program and \cvf transitions. 
Our focus in this paper was on stabilizing programs because they are naturally \cvf-tolerant and there is a vast literature that provides stabilizing algorithms. However, another future work is in developing algorithms that tolerate \cvf{s} without being stabilizing.

We observe that for the token ring program that on a logarithmic scale, the slope of the (approximation) curve of \cvf{s} against rank effect decreases, except for one anomaly. It means that $\dfrac{\Delta\text{ cvf occurence}}{\Delta\text{ Rank change value}}$
keeps decreasing as the order of the input graph increases. We do not have the same observations for coloring and maximal matching programs. 
Hence, one of the future works is to study and possibly determine how the \cvf distribution changes with the size of the input. 



\bibliography{main.bib}
\bibliographystyle{acm}

\newpage
\appendix




\section{Appendix: Case study programs}
\label{sec:case-study}

In this section, we present three case study programs used in our evaluation: token ring, graph coloring, and maximal matching.

\textbf{Token ring program:} The 3-state token ring program is one of the classic self-stabilizing program introduced by Dijkstra \cite{EDW426}. In this program, we have $n$ processes, $0, 1, \dots, n-1$, organized in a ring. Each process $j$ has a variable $x.j$ in the domain $\{0, 1, 2\}$. Process 0 checks if $x.0+1$ equals $x.1$. If so, process 0 decrements the value of $x.0$. (All operations of this token ring program are in modulo 3 arithmetic). Process $n-1$ checks if $x.(n-2)=x.0$ and $x.(n-2)+1 \neq x.(n-1)$. If so, process $n-1$ copies the value of $x.(n-2)$. Other processes check if their $x$ value plus 1 is equals to their left (or right, respectively) neighbor. If that is true, it copies the $x$ value of the left (or right, respectively) neighbor. The actions of the processes are as follows:

$\text{At 0}: (x.0 + 1) \mod 3 == x.1 \longrightarrow x.0 = (x.0 - 1) \mod 3;$

$\text{At } n-1: (x.0 == x.(n-2)) \wedge ((x.(n-2)+1) \mod 3 \neq x.(n-1)) \longrightarrow x.(n-1) = x.(n-2);$

$\text{At $j$ where $1 \leq j \leq n-2$ : } 
(x.j +1) \mod 3 == x.(j-1) \longrightarrow x.j = x.(j-1);$ 

$\text{At $j$ where $1 \leq j \leq n-2$ : }
(x.j +1) \mod 3 == x.(j+1) \longrightarrow x.j = x.(j+1);$

\textbf{Graph coloring program:} We use the self-stabilizing graph coloring program by Gradinariu and Tixeuil~\cite{GT2000OPODIS.short}. This program guarantees to use no more than $\Delta+1$ colors where $\Delta$ is the maximum degree of the graph.
In this program, each process $j$ is associated with a color $c.j$. Each process has one action. It checks the color of its neighbors. If $c.j$ is the same as $c.k$ for some neighbor $k$ then $c.j$ is changed to a color not present in the neighborhood.
Thus, the action at process $j$ is as follows. 

{\small $c.j \in \{c.k|k \in Nb.j\} \longrightarrow c.j = c$, where $c = \min\{l \in \mathbb{N} | l \not\in \{c.k|k \in Nb.j\} \}$.}

\textbf{Maximal-matching program: } We use the self-stabilizing maximal matching program by Manne et al.~\cite{MMPT2009TCS.short}. In this program, each process $j$ is associated with an integer $p.j$. 
The value of $p.j$ indicates the process with which $j$ proposes to be matched. If $p.j == null$ then $j$ currently does not propose to any process. If two processes $i$ and $j$ propose with each other, i.e. $p.i == j$ and $p.j == i$, then we say the two processes are married. The marital status of a process $j$ is determined by evaluating the predicate $PRmarried(j) \equiv \exists i \in Nb.j: (p.i = j) \wedge (p.j = i)$. A free process $j$ will not try to get matched with a married neighbor. However, $j$ is not able to evaluate the predicate $PRmarried()$ of its neighbors because the evaluation requires information about the neighbors of $j$'s neighbors. To help processes communicate their marital status with their neighbors, each process $j$ is also associated with a Boolean variable $m.j$ indicating whether process $j$ has been married or not. It is possible that the value of $m.j$ is not consistent with $PRmarried(j)$. In this case, $m.j$ has to be updated to be consistent.

When the value of $m.j$ is consistent with $PRmarried(j$, the maximal matching algorithm at $j$ works as follows. If $j$ currently does not propose to anyone and there is a neighbor who proposes to $j$, then $j$ proposes back to that neighbor. This action will match 2 processes together. If $j$ currently does not propose to anyone and none of the neighbors proposes to $j$ then $j$ will propose to a neighbor who has free marital status and smaller ID than $j$ and currently does not propose to anyone. If there are multiple such neighbors, $j$ will choose the one with the smallest ID. This action will let free processes find a new partner. If $j$ proposes to a neighbor $i$ but $i$ does not propose to $j$ and $i$ is married then $j$ will cancel its proposal. This action will protect existing matches. If $j$ proposes to a neighbor $i$ but $i$ does not propose to $j$ and the ID of $i$ is greater than the ID of $j$ then $j$ will cancel its proposal. This action is to break the symmetry. The actions at process $j$ are as follows.

$m.j \neq PRmarried(j) \longrightarrow m.j = PRmarried(j);$

$m.j == PRmarried(j) \wedge p.j == null \wedge \exists i \in Nb.j: p.i == j \longrightarrow p.j = i;$

$m.j == PRmarried(j) \wedge p.j == null \wedge \forall i \in Nb.j: p.i \neq j \wedge \exists k \in Nb.j: (p.k == null \wedge k < j \wedge \neg m.k) \longrightarrow p.j = max\{k \in Nb.j: (p.k == null \wedge k < j \wedge \neg m.k) \};$

$m.j == PRmarried(j) \wedge p.j == i \neq null \wedge p.i \neq j \wedge (m.i \vee j \leq i) \longrightarrow p.j = null;$


\section{Defining \mrank and \arank}
\label{sec:define-ranks}


The \mrank of each state is computed as follows: First, we set the rank of all legitimate states to $0$. Subsequently, we find a state $s_0$ such that $rank(s_0)$ is still unknown but the rank of all successors of $s_0$ is known. Note that successors of $s_0$ are $\{ s_1 | (s_0, s_1)$ is a transition of the given program $\}$. Thus, $rank(s_0)$ is set to $max(rank(s_1) + 1| (s_0, s_1)$ is a transition of the given program.)

To compute average rank, we use a similar approach. Each state $s$ is associated with $pathCount$ and $totalPathLength$ where $pathCount$ is the number of paths from $s$ to the first state in the invariant and $totalPahtLength$ is the sum of the lengths of those paths. The average rank is computed as $\arank = \frac{totalPathLength}{pathCount}$. For states in the invariant, we set $pathCount = 1$ and $totalPathLength = 0$. Subsequently, we find a state $s_0$ whose rank is unknown but the ranks of all its successor is known. Then we update $totalPathLength$ and $pathCount$ for $s_0$. Specifically, let $succ$ denote the set of all successors of $s_0$ by program transitions and $|succ|$ is the number of successors of $s_0$. Then $pathCount(s_0) = \sum\limits_{s_1 \in succ} pathCount(s_1)$ and $totalPathLength(s_0) = |succ| + \sum\limits_{s_1 \in succ} totalPathLength(s_1)$.


\end{document}